# Matrix genetics, part 2: the degeneracy of the genetic code and the octave algebra with two quasi-real units (the genetic octave Yin-Yang-algebra)

Sergey V. Petoukhov

Department of Biomechanics, Mechanical Engineering Research Institute of the
Russian Academy of Sciences
petoukhov@hotmail.com, petoukhov@imash.ru, http://symmetry.hu/isabm/petoukhov.html

**Abstract**. Algebraic properties of the genetic code are analyzed. The investigations of the genetic code on the basis of matrix approaches ("matrix genetics") are described. The degeneracy of the vertebrate mitochondria genetic code is reflected in the black-and-white mosaic of the (8*8)-matrix of 64 triplets, 20 amino acids and stop-signals. This mosaic genetic matrix is connected with the matrix form of presentation of the special 8-dimensional Yin-Yang-algebra and of its particular 4-dimensional case. The special algorithm, which is based on features of genetic molecules, exists to transform the mosaic genomatrix into the matrices of these algebras. Two new numeric systems are defined by these 8-dimensional and 4-dimensional algebras: genetic Yin-Yang-octaves and genetic tetrions. Their comparison with quaternions by Hamilton is presented. Elements of new "genovector calculation" and ideas of "genetic mechanics" are discussed. These algebras are considered as models of the genetic code and as its possible pre-code basis. They are related with binary oppositions of the Yin-Yang type and they give new opportunities to investigate evolution of the genetic code. The revealed fact of the relation between the genetic code and these genetic algebras is discussed in connection with the idea by Pythagoras: "All things are numbers". Simultaneously these genetic algebras can be utilized as the algebras of genetic operators in biological organisms. The described results are related with the problem of algebraization of bioinformatics. They take attention to the question: what is life from the viewpoint of algebra?

## 1 Introduction

This article is devoted to algebraic properties of molecular systems of the genetic code in their matrix representations. The initial approaches for investigations of genetic code systems from the viewpoint of matrix approaches were described in our previous publications [Petoukhov, 2001-2008]. These investigations are generalized under the name "matrix genetics". They are connected closely with matrix forms of digital signal processing in computers.

From an information viewpoint, biological organisms are informational essences. They receive genetic information from their ancestors and transmit it to descendants. A conception of informational nature of living organisms is reflected in the words: "If you want to understand life, don't think about vibrant, throbbing dels and oozes, think about information technology" [Dawkins, 1991]. Or another citation of a similar direction of thoughts: "Notions of "information" or "valuable information" are not utilized in physics of non-biological nature because they are not needed there. On the contrary, in biology notions "information" and especially "valuable information" are main ones; understanding and description of phenomena in biological nature are impossible without these notions. A specificity of "living matter" lies in them" [Chernavskiy, 2000]. Bioinformatics can give deeper knowledge in the questions what is life and why life exists. A development of theoretical biology needs in appropriate mathematical models of structural ensembles of genetic elements. The effective matrix approach for such models is proposed below.

## 2 The genetic octave matrix as the matrix form of presentation of the octave algebra

Algebras of complex and hypercomplex numbers $x_0*\mathbf{1}+x_1*\mathbf{i}_1+\ldots+x_k*\mathbf{i}_k$ are well-known (the usual definition of the term "algebra over a field P" is given in the Appendix B). It is known also that complex and hypercomplex numbers have not only vector forms of their presentations, but also matrix forms of their presentation. For example complex numbers $z = x*\mathbf{1}+y*\mathbf{i}$ (where $\mathbf{1}$ is the real unit and $\mathbf{i}^2 = -1$ is the imaginary unit) possess the following matrix form of their presentation:

$$z = x*\mathbf{1} + y*\mathbf{i} = x* \begin{vmatrix} 1 & 0 \\ 0 & 1 \end{vmatrix} + y* \begin{vmatrix} 0 & 1 \\ -1 & 0 \end{vmatrix} = \begin{vmatrix} x & y \\ -y & x \end{vmatrix} \quad (1)$$

By the way, complex numbers are utilized in computers in this matrix form. The following table of multiplication of the basic matrix elements $1=[1\ 0;\ 0\ 1]$ and $\mathbf{i} = [0\ 1;\ -1\ 0]$ for the algebra of complex numbers exists:

|   | 1 | i  |
|---|---|----|
| 1 | 1 | i  |
| i | i | -1 |

(2)

Our initial idea is concluded in interpretation of the genetic matrices as matrix forms of presentation of special algebras (or systems of multidimensional numbers) on the basis of molecular features of the letters C, A, U/T, G of the genetic alphabet. Let us apply this idea to the genetic matrix $P^{CAUG}_{123}{}^{(3)}$ (or $P^{(3)}=[C\ A;\ U\ G]^{(3)}$), which was described in our previous articles [Petoukhov, arXiv:0802.3366; arXiv:0803.0888] for the case of the vertebrate mitochondria genetic code (Figure 1). This genomatrix has 32 "black" triplets and 32 "white" triplets disposed in matrix cells of appropriate colors (see details in [Petoukhov, arXiv:0802.3366; arXiv:0803.0888]). Both quadrants along each of two diagonals of this genomatrix possess identical mosaics. All triplets in two quadrants along the main diagonal begin with the letters C and G, which possess 3 hydrogen bonds in their complementary pair C-G. All triplets in two quadrants along the second diagonal begin with the letters A and U, which possess 2 hydrogen bonds in their complementary pair A-U.

**The phenomenological "alphabetic" rule # 1** exists for the matrix disposition of triplets of black and white colors:
- In the set of 32 triplets, the first letter of which has 3 hydrogen bonds (the letters C and G), the white triplets are those ones, the second position of which is occupied by the letter A (that is by the purine with 2 hydrogen bonds); the other triplets of this set are black triplets;
- In the set of 32 triplets, the first letter of which has 2 hydrogen bonds (the letters A and U), the black triplets are those ones, the second position of which is occupied by the letter C (that is by the pyrimidine with 3 hydrogen bonds); the other triplets of this set are white triplets.

For example in accordance with this rule the triplet CAG is the white triplet because its first letter C has 3 hydrogen bonds, and its second position is occupied by the letter A. It is obvious that the first part of this rule, which utilizes molecular features of the genetic letters C, A, G, U, is related to triplets of two quadrants along the main matrix diagonal, and that the second part of this rule is related to triplets of two quadrants along the second diagonal.

We should note the inessential modification in numeration of the columns and the rows in this article in comparison with our previous article [Petoukhov, arXiv:0803.0888], where the decreasing sequence 111 (7), 110 (6), 101 (5), 100 (4), 011 (3), 010 (2), 001 (1), 000 (0) in the

genomatrix $P^{CAUG}_{123}{}^{(3)}$ was utilized. It was made to demonstrate the coincidence with the famous table of 64 hexagrams of the ancient Chinese book "I Ching", which possessed this decreasing sequence. Now we number the columns and the rows of this genomatrix on Figure 1 by the ascending sequence 000 (0), 001 (1), 010 (2), 011 (3), 100 (4), 101 (5), 110 (6), 111 (7), which is more traditional for matrix analysis and for the theory of digital signal processing. The columns and the rows are numbered by this ascending sequence on the basis of their triplets algorithmically, if we change the correspondence between binary symbols (0 and 1) and the genetic letters in the two first genetic sub-alphabets [Petoukhov, arXiv:0803.0888] by assuming the following:

| $P^{CAUG}_{123}{}^{(3)}$: | 000 (0) | 001 (1) | 010 (2) | 011 (3) | 100 (4) | 101 (5) | 110 (6) | 111 (7) |
|---|---|---|---|---|---|---|---|---|
| **000 (0)** | CCC Pro | CCA Pro | CAC His | CAA Gln | ACC Thr | ACA Thr | AAC Asn | AAA Lys |
| **001 (1)** | CCU Pro | CCG Pro | CAU His | CAG Gln | ACU Thr | ACG Thr | AAU Asn | AAG Lys |
| **010 (2)** | CUC Leu | CUA Leu | CGC Arg | CGA Arg | AUC Ile | AUA Met | AGC Ser | AGA Stop |
| **011 (3)** | CUU Leu | CUG Leu | CGU Arg | CGG Arg | AUU Ile | AUG Met | AGU Ser | AGG Stop |
| **100 (4)** | UCC Ser | UCA Ser | UAC Tyr | UAA Stop | GCC Ala | GCA Ala | GAC Asp | GAA Glu |
| **101 (5)** | UCU Ser | UCG Ser | UAU Tyr | UAG Stop | GCU Ala | GCG Ala | GAU Asp | GAG Glu |
| **110 (6)** | UUC Phe | UUA Leu | UGC Cys | UGA Trp | GUC Val | GUA Val | GGC Gly | GGA Gly |
| **111 (7)** | UUU Phe | UUG Leu | UGU Cys | UGG Trp | GUU Val | GUG Val | GGU Gly | GGG Gly |

Figure 1. The genomatrix $P^{CAUG}_{123}{}^{(3)}$, each cell of which has a triplet and an amino acid (or stop-signal) coded by this triplet. The black-and-white mosaic presents a specificity of the degeneracy of the vertebrate mitochondria genetic code (from [Petoukhov, arXiv:0803.0888]).

| | | 000 (0) | 001 (1) | 010 (2) | 011 (3) | 100 (4) | 101 (5) | 110 (6) | 111 (7) |
|---|---|---|---|---|---|---|---|---|---|
| | 000 (0) | $x_0$ | $x_1$ | $-x_2$ | $-x_3$ | $x_4$ | $x_5$ | $-x_6$ | $-x_7$ |
| | 001 (1) | $x_0$ | $x_1$ | $-x_2$ | $-x_3$ | $x_4$ | $x_5$ | $-x_6$ | $-x_7$ |
| | 010 (2) | $x_2$ | $x_3$ | $x_0$ | $x_1$ | $-x_6$ | $-x_7$ | $-x_4$ | $-x_5$ |
| $YY_8 =$ | 011 (3) | $x_2$ | $x_3$ | $x_0$ | $x_1$ | $-x_6$ | $-x_7$ | $-x_4$ | $-x_5$ |
| | 100 (4) | $x_4$ | $x_5$ | $-x_6$ | $-x_7$ | $x_0$ | $x_1$ | $-x_2$ | $-x_3$ |
| | 101 (5) | $x_4$ | $x_5$ | $-x_6$ | $-x_7$ | $x_0$ | $x_1$ | $-x_2$ | $-x_3$ |
| | 110 (6) | $-x_6$ | $-x_7$ | $-x_4$ | $-x_5$ | $x_2$ | $x_3$ | $x_0$ | $x_1$ |
| | 111 (7) | $-x_6$ | $-x_7$ | $-x_4$ | $-x_5$ | $x_2$ | $x_3$ | $x_0$ | $x_1$ |

Figure 2. The matrix $YY_8$ as the matrix form of presentation of the genetic octave algebra with two quasi-real units (the genetic octave Yin-Yang-algebra). The black cells contain coordinates with the sign „+" and the white cells contain coordinates with the sign „-".

- the first genetic sub-alphabet, that defines the binary number of the columns, presents each pyrimidine (C and U/T) by the symbol 0, and presents each purine (A and G) by the symbol 1;
- the second genetic sub-alphabet, that defines the binary number of the rows, presents each letter with the amino-mutating property (C and A) by the symbol *0*, and presents each letter without such property (G and U/T) by the symbol *1*.

Taking into account the molecular characteristics of the nitrogenous bases C, A, U/T, G of the genetic alphabet, one can reform the genomatrix $P^{CAUG}_{123}{}^{(3)}$ into the new matrix $YY_8$ algorithmically (Figure 2). The mosaic of the disposition of signs "+" (it occupies the black matrix cells) and "-" (it occupies the white matrix cells) in matrix $YY_8$ is identical to the mosaic of the genomatrix $P^{CAUG}_{123}{}^{(3)}$ (Figure 1). Below we shall list the structural analogies between these matrices $P^{CAUG}_{123}{}^{(3)}$ and $YY_8$ and demonstrate that this matrix $YY_8$ is the matrix representation of the octave algebra with two quasi-real units. But initially we pay attention to the "alphabetic" algorithm of digitization of 64 triplets, which gives the matrix $YY_8$ from the genomatrix $P^{CAUG}_{123}{}^{(3)}$.

**2.1 The alphabetic algorithm of the Yin-Yang-digitization of 64 triplets**

This algorithm is based on utilizing the two binary-oppositional attributes of the genetic letters C, A, U/T, G: "purine or pyrimidine" and "2 or 3" hydrogen bonds. It uses also the famous thesis of molecular genetics that different positions inside triplets have different code meanings. For example the article [Konopelchenko, Rumer, 1975] has described that two first positions of each triplet form "the root of the codon" and that they differ drastically from the third position by their essence and by their special role. Because of this "alphabetic" algorithm, the transformation of the genomatrix $P^{CAUG}_{123}{}^{(3)}$ into the matrix $YY_8$ is not an abstract and arbitrary action at all, but such transformation can be utilized by biocomputer systems of organisms practically.

The alphabetic algorithm of the Yin-Yang-digitization defines the special scheme of reading of each triplet: the first two positions of the triplet are read by genetic systems from the viewpoint of one attribute (the attribute of "2 or 3" hydrogen bonds) and the third position of the triplet is read from the viewpoint of another attribute (the attribute of "purine or pyrimidine"). The algorithm consists of three parts, where the first two parts define the generalized numeric symbol of each triplet and the third part defines its sign "+" or "-":
1. Two first positions of each triplet is read from the viewpoint of the binary-oppositional attribute "2 or 3 hydrogen bonds" of the genetic letters: each letter from the complementary pair C and G is interpreted as a real number α (for instance, α=3 because C and G have 3 hydrogen bonds each), and each letter from the second complementary pair A and U/T is interpreted as a real number β (for instance, β =2 because A and U/T have 2 hydrogen bonds each).
2. The third position of each triplet is read from the viewpoint of the another binary-oppositional attribute "purine or pyrimidine": each pyrimidine C or U/T is interpreted as a real number γ (for instance γ=1 because each pyrimidine contains one molecular ring), and each purine A or G is interpreted as a real number δ (for instance, δ=2 because each purine contains two molecular rings).
3. The generalized numeric symbol of each black (white) triplet has a sign "+" ("-" correspondingly); the definition of the black triplets and the white triplets was made above in the rule # 1 on the basis of molecular properties of the genetic letters also.

For example the triplet CAG receives the generalized numeric symbol "-αβδ" by this algorithm because its first letter C is symbolized as "α", the second letter is symbolized as "β" and the third letter G is symbolized as "δ". The sign "-" appears because CAG is the white triplet in accordance with the "alphabetic" rule # 1. The described algorithm can be considered as the algorithm of special conversion by means of which the four genetic letters are substituted for four real numbers α, β, γ, δ, and each triplet appears in the form of the chain (or the ensemble) of these real numbers with the appropriate sign. One can say that new alphabet of the four symbols α, β, γ, δ appears (see the section 6 about the numeric system of tetrions as the genetic pre-code). Figure 3 illustrates the details of such algorithmic conversion of the genomatrix $P^{CAUG}_{123}(3)$ into the matrix $YY_8$, where the 8 variants of the 3-digit chains take place as components of the matrix $YY_8$: ααγ=$x_0$, ααδ=$x_1$, αβγ=$x_2$, αβδ=$x_3$, βαγ=$x_4$, βαδ=$x_5$, ββγ=$x_6$, ββδ=$x_7$. We shall name these matrix components $x_0, x_1, \ldots, x_7$, which are real numbers, as "YY-coordinates".

|   | 000 (0) | 001 (1) | 010 (2) | 011 (3) | 100 (4) | 101 (5) | 110 (6) | 111 (7) |
|---|---|---|---|---|---|---|---|---|
| **000 (0)** | CCC<br>ααγ<br>$x_0$ | CCA<br>ααδ<br>$x_1$ | CAC<br>-αβγ<br>-$x_2$ | CAA<br>-αβδ<br>-$x_3$ | ACC<br>βαγ<br>$x_4$ | ACA<br>βαδ<br>$x_5$ | AAC<br>-ββγ<br>-$x_6$ | AAA<br>-ββδ<br>-$x_7$ |
| **001 (1)** | CCU<br>ααγ<br>$x_0$ | CCG<br>ααδ<br>$x_1$ | CAU<br>-αβγ<br>-$x_2$ | CAG<br>-αβδ<br>-$x_3$ | ACU<br>βαγ<br>$x_4$ | ACG<br>βαδ<br>$x_5$ | AAU<br>-ββγ<br>-$x_6$ | AAG<br>-ββδ<br>-$x_7$ |
| **010 (2)** | CUC<br>αβγ<br>$x_2$ | CUA<br>αβδ<br>$x_3$ | CGC<br>ααγ<br>$x_0$ | CGA<br>ααδ<br>$x_1$ | AUC<br>-ββγ<br>-$x_6$ | AUA<br>-ββδ<br>-$x_7$ | AGC<br>-βαγ<br>-$x_4$ | AGA<br>-βαδ<br>-$x_5$ |
| **011 (3)** | CUU<br>αβγ<br>$x_2$ | CUG<br>αβδ<br>$x_3$ | CGU<br>ααγ<br>$x_0$ | CGG<br>ααδ<br>$x_1$ | AUU<br>-ββγ<br>-$x_6$ | AUG<br>-ββδ<br>-$x_7$ | AGU<br>-βαγ<br>-$x_4$ | AGG<br>-βαδ<br>-$x_5$ |
| **100 (4)** | UCC<br>βαγ<br>$x_4$ | UCA<br>βαδ<br>$x_5$ | UAC<br>-ββγ<br>-$x_6$ | UAA<br>-ββδ<br>-$x_7$ | GCC<br>ααγ<br>$x_0$ | GCA<br>ααδ<br>$x_1$ | GAC<br>-αβγ<br>-$x_2$ | GAA<br>-αβδ<br>-$x_3$ |
| **101 (5)** | UCU<br>βαγ<br>$x_4$ | UCG<br>βαδ<br>$x_5$ | UAU<br>-ββγ<br>-$x_6$ | UAG<br>-ββδ<br>-$x_7$ | GCU<br>ααγ<br>$x_0$ | GCG<br>ααδ<br>$x_1$ | GAU<br>-αβγ<br>-$x_2$ | GAG<br>-αβδ<br>-$x_3$ |
| **110 (6)** | UUC<br>-ββγ<br>-$x_6$ | UUA<br>-ββδ<br>-$x_7$ | UGC<br>-βαγ<br>-$x_4$ | UGA<br>-βαδ<br>-$x_5$ | GUC<br>αβγ<br>$x_2$ | GUA<br>αβδ<br>$x_3$ | GGC<br>ααγ<br>$x_0$ | GGA<br>ααδ<br>$x_1$ |
| **111 (7)** | UUU<br>-ββγ<br>-$x_6$ | UUG<br>-ββδ<br>-$x_7$ | UGU<br>-βαγ<br>-$x_4$ | UGG<br>-βαδ<br>-$x_5$ | GUU<br>αβγ<br>$x_2$ | GUG<br>αβδ<br>$x_3$ | GGU<br>ααγ<br>$x_0$ | GGG<br>ααδ<br>$x_1$ |

Figure 3. The result of the algorithmic conferment of 64 triplets to numeric coordinates $x_0, x_1, \ldots, x_7$, which are based on the four real numbers α, β, γ, δ.

In the section 2.3 we will describe the structural analogies between sets of elements of the genomatrix $P^{CAUG}_{123}(3)$ and of the matrix $YY_8$. But now we will pay attention to algebraic properties of the matrix $YY_8$.

### 2.2 The Yin-Yang-genomatrix $YY_8$ as the element of the octave Yin-Yang-algebra

It is quite unexpectedly that this new matrix $YY_8$ (Figure 2), which is constructed algorithmically from the genomatrix $P^{CAUG}_{123}(3)$, presents the unusual algebra with such set of basic elements which contains two quasi-real units and does not contain the real unit at all. Really the matrix $YY_8$ with its 8 coordinates $x_0, x_1, \ldots, x_7$ can be represented as the sum of the 8 matrices, each of which contains only one of these coordinates (Figure 4). Let us symbolize any matrix, which is multiplied there by any of YY-coordinates $x_0, x_2, x_4, x_6$ with even indexes, by the symbol $f_k$ (where "f" is the first letter of the word "female" and k=0, 2, 4, 6). We will mark these matrices $f_k$ and their coordinates $x_0, x_2, x_4, x_6$ by pink color. And let us symbolize any matrix, which is multiplied there by any of YY-coordinates $x_1, x_3, x_5, x_7$ with odd indexes, by the symbol $m_p$ (where "m" is the first letter of the word "male" and p=1, 3, 5, 7). We will mark these matrices

$\mathbf{m_p}$ and their coordinates $x_1, x_3, x_5, x_7$ by blue color. In this case one can present the matrix $YY_8$ in the form (3).

$$YY_8 = x_0*\mathbf{f_0}+x_1*\mathbf{m_1}+x_2*\mathbf{f_2}+x_3*\mathbf{m_3}+x_4*\mathbf{f_4}+x_5*\mathbf{m_5}+x_6*\mathbf{f_6}+x_7*\mathbf{m_7} \qquad (3)$$

$YY_8 = x_0*\begin{pmatrix} 1 & 0 & 0 & 0 & 0 & 0 & 0 & 0 \\ 1 & 0 & 0 & 0 & 0 & 0 & 0 & 0 \\ 0 & 0 & 1 & 0 & 0 & 0 & 0 & 0 \\ 0 & 0 & 1 & 0 & 0 & 0 & 0 & 0 \\ 0 & 0 & 0 & 0 & 1 & 0 & 0 & 0 \\ 0 & 0 & 0 & 0 & 1 & 0 & 0 & 0 \\ 0 & 0 & 0 & 0 & 0 & 0 & 1 & 0 \\ 0 & 0 & 0 & 0 & 0 & 0 & 1 & 0 \end{pmatrix} + x_1*\begin{pmatrix} 0 & 1 & 0 & 0 & 0 & 0 & 0 & 0 \\ 0 & 1 & 0 & 0 & 0 & 0 & 0 & 0 \\ 0 & 0 & 0 & 1 & 0 & 0 & 0 & 0 \\ 0 & 0 & 0 & 1 & 0 & 0 & 0 & 0 \\ 0 & 0 & 0 & 0 & 0 & 1 & 0 & 0 \\ 0 & 0 & 0 & 0 & 0 & 1 & 0 & 0 \\ 0 & 0 & 0 & 0 & 0 & 0 & 0 & 1 \\ 0 & 0 & 0 & 0 & 0 & 0 & 0 & 1 \end{pmatrix} +$

$+ x_2*\begin{pmatrix} 0 & 0 & -1 & 0 & 0 & 0 & 0 & 0 \\ 0 & 0 & -1 & 0 & 0 & 0 & 0 & 0 \\ 1 & 0 & 0 & 0 & 0 & 0 & 0 & 0 \\ 1 & 0 & 0 & 0 & 0 & 0 & 0 & 0 \\ 0 & 0 & 0 & 0 & 0 & 0 & -1 & 0 \\ 0 & 0 & 0 & 0 & 0 & 0 & -1 & 0 \\ 0 & 0 & 0 & 0 & 1 & 0 & 0 & 0 \\ 0 & 0 & 0 & 0 & 1 & 0 & 0 & 0 \end{pmatrix} + x_3*\begin{pmatrix} 0 & 0 & 0 & -1 & 0 & 0 & 0 & 0 \\ 0 & 0 & 0 & -1 & 0 & 0 & 0 & 0 \\ 0 & 1 & 0 & 0 & 0 & 0 & 0 & 0 \\ 0 & 1 & 0 & 0 & 0 & 0 & 0 & 0 \\ 0 & 0 & 0 & 0 & 0 & 0 & 0 & -1 \\ 0 & 0 & 0 & 0 & 0 & 0 & 0 & -1 \\ 0 & 0 & 0 & 0 & 0 & 1 & 0 & 0 \\ 0 & 0 & 0 & 0 & 0 & 1 & 0 & 0 \end{pmatrix} +$

$+ x_4*\begin{pmatrix} 0 & 0 & 0 & 0 & 1 & 0 & 0 & 0 \\ 0 & 0 & 0 & 0 & 1 & 0 & 0 & 0 \\ 0 & 0 & 0 & 0 & 0 & 0 & -1 & 0 \\ 0 & 0 & 0 & 0 & 0 & 0 & -1 & 0 \\ 1 & 0 & 0 & 0 & 0 & 0 & 0 & 0 \\ 1 & 0 & 0 & 0 & 0 & 0 & 0 & 0 \\ 0 & 0 & -1 & 0 & 0 & 0 & 0 & 0 \\ 0 & 0 & -1 & 0 & 0 & 0 & 0 & 0 \end{pmatrix} + x_5*\begin{pmatrix} 0 & 0 & 0 & 0 & 0 & 1 & 0 & 0 \\ 0 & 0 & 0 & 0 & 0 & 1 & 0 & 0 \\ 0 & 0 & 0 & 0 & 0 & 0 & 0 & -1 \\ 0 & 0 & 0 & 0 & 0 & 0 & 0 & -1 \\ 0 & 1 & 0 & 0 & 0 & 0 & 0 & 0 \\ 0 & 1 & 0 & 0 & 0 & 0 & 0 & 0 \\ 0 & 0 & 0 & -1 & 0 & 0 & 0 & 0 \\ 0 & 0 & 0 & -1 & 0 & 0 & 0 & 0 \end{pmatrix} +$

$+ x_6*\begin{pmatrix} 0 & 0 & 0 & 0 & 0 & 0 & -1 & 0 \\ 0 & 0 & 0 & 0 & 0 & 0 & -1 & 0 \\ 0 & 0 & 0 & 0 & -1 & 0 & 0 & 0 \\ 0 & 0 & 0 & 0 & -1 & 0 & 0 & 0 \\ 0 & 0 & -1 & 0 & 0 & 0 & 0 & 0 \\ 0 & 0 & -1 & 0 & 0 & 0 & 0 & 0 \\ -1 & 0 & 0 & 0 & 0 & 0 & 0 & 0 \\ -1 & 0 & 0 & 0 & 0 & 0 & 0 & 0 \end{pmatrix} + x_7*\begin{pmatrix} 0 & 0 & 0 & 0 & 0 & 0 & 0 & -1 \\ 0 & 0 & 0 & 0 & 0 & 0 & 0 & -1 \\ 0 & 0 & 0 & 0 & 0 & -1 & 0 & 0 \\ 0 & 0 & 0 & 0 & 0 & -1 & 0 & 0 \\ 0 & 0 & 0 & -1 & 0 & 0 & 0 & 0 \\ 0 & 0 & 0 & -1 & 0 & 0 & 0 & 0 \\ 0 & -1 & 0 & 0 & 0 & 0 & 0 & 0 \\ 0 & -1 & 0 & 0 & 0 & 0 & 0 & 0 \end{pmatrix}$

Figure 4. The presentation of the matrix $YY_8$ as the sum of the 8 matrices.

The important fact is that the set of these 8 matrices $\mathbf{f_0}, \mathbf{m_1}, \mathbf{f_2}, \mathbf{m_3}, \mathbf{f_4}, \mathbf{m_5}, \mathbf{f_6}, \mathbf{m_7}$ forms the closed set relative to multiplications: a multiplication between any two matrices from this set generates a matrix from this set again. The table on Figure 5 presents the results of multiplications among these 8 matrices. It is known that such multiplication tables define appropriate algebras over a field (see Appendix B). Correspondingly the table on Figure 5 defines the genetic octave algebra $YY_8$. Multiplication of any two members of the octave algebra $YY_8$ generates a new member of the same algebra. This situation is similar to the situation of real numbers (or of complex numbers, or of hypercomplex numbers) when multiplication of any two members of the numeric system generates a new member of the same numerical system. In other words, we receive new numerical system of $YY_8$ octaves (3) from the natural structure of the genetic code.

|       | $f_0$ | $m_1$ | $f_2$  | $m_3$  | $f_4$  | $m_5$  | $f_6$ | $m_7$ |
|-------|-------|-------|--------|--------|--------|--------|-------|-------|
| $f_0$ | $f_0$ | $m_1$ | $f_2$  | $m_3$  | $f_4$  | $m_5$  | $f_6$ | $m_7$ |
| $m_1$ | $f_0$ | $m_1$ | $f_2$  | $m_3$  | $f_4$  | $m_5$  | $f_6$ | $m_7$ |
| $f_2$ | $f_2$ | $m_3$ | $-f_0$ | $-m_1$ | $-f_6$ | $-m_7$ | $f_4$ | $m_5$ |
| $m_3$ | $f_2$ | $m_3$ | $-f_0$ | $-m_1$ | $-f_6$ | $-m_7$ | $f_4$ | $m_5$ |
| $f_4$ | $f_4$ | $m_5$ | $f_6$  | $m_7$  | $f_0$  | $m_1$  | $f_2$ | $m_3$ |
| $m_5$ | $f_4$ | $m_5$ | $f_6$  | $m_7$  | $f_0$  | $m_1$  | $f_2$ | $m_3$ |
| $f_6$ | $f_6$ | $m_7$ | $-f_4$ | $-m_5$ | $-f_2$ | $-m_3$ | $f_0$ | $m_1$ |
| $m_7$ | $f_6$ | $m_7$ | $-f_4$ | $-m_5$ | $-f_2$ | $-m_3$ | $f_0$ | $m_1$ |

Figure 5. The multiplication table of the Yin-Yang-algebra $YY_8$ for the case of $P^{CAUG}_{123}{}^{(3)}$.

In accordance with its multiplication table (Figure 5), the algebra of this new numeric system contains two quasi-real units $f_0$ и $m_1$ in the set of its 8 basic matrices and it does not contain the real unit **1** at all. Really the set of basic matrices $f_0$, $m_1$, $f_2$, $m_3$, $f_4$, $m_5$, $f_6$, $m_7$ is divided into two equal sub-sets by attributes of their squares. The first sub-set contains $f_0$, $f_2$, $f_4$, $f_6$. The squares of these basic matrices $f_0$, $f_2$, $f_4$, $f_6$ are equal to $\pm f_0$ always. We name this sub-set the $f_0$-sub-set of the basic matrices (or of the basic elements). The second sub-set contains $m_1$, $m_3$, $m_5$, $m_7$. We name this sub-set the $m_1$-sub-set of the basic elements. The squares of these basic matrices $m_1$, $m_3$, $m_5$, $m_7$ are equal to $\pm m_1$ always (see elements on the main diagonal of the multiplication table on Figure 5). The YY-coordinates $x_0$, $x_2$, $x_4$, $x_6$, which are connected with the basic elements $f_0$, $f_2$, $f_4$, $f_6$, form the $f_0$-sub-set of the eight YY-coordinates correspondingly. The YY-coordinates $x_1$, $x_3$, $x_5$, $x_7$, which are connected with the basic elements $m_1$, $m_3$, $m_5$, $m_7$, form the $m_1$-sub-set.

The initial basic element $f_0$ possesses all properties of the real unit in relation to each member of the $f_0$-sub-set of the basic elements: $f_0^2 = f_0$, $f_0*f_2=f_2*f_0=f_2$, $f_0*f_4=f_4*f_0=f_4$, $f_0*f_6=f_6*f_0=f_6$. But the element $f_0$ loses the commutative property of the real unit in relation to the members of the $m_1$-sub-set of the basic elements: $f_0*m_p \neq m_p*f_0$, where p = 1, 3, 5, 7. In this reason the element $f_0$ is named the quasi-real unit from the $f_0$-sub-set of the basic elements.

By analogy the basic element $m_1$ possesses all properties of the real unit in relation to each member of the second sub-set of the basic elements $m_1$, $m_3$, $m_5$, $m_7$: $m_1^2=m_1$, $m_1*m_3=m_3*m_1=m_3$, $m_1*m_5=m_5*m_1=m_5$, $m_1*m_7=m_7*m_1=m_7$. But the element $m_1$ loses the commutative property of the real unit in relation to members of the $f_0$-sub-set: $m_1*f_k \neq f_k*m_1$, where k = 0, 2, 4, 6. In this reason the element $m_1$ is named the quasi-real unit from the $m_1$-sub-set of the basic elements.

Let us pay attention to the unexpected circumstance. All members of the $f_0$-sub-set of the basic elements $f_0$, $f_2$, $f_4$, $f_6$ and of their coordinates $x_0$, $x_2$, $x_4$, $x_6$ have the even indexes 0, 2, 4, 6 (zero is considered as even number here). And they occupy the columns with the even numbers 0, 2, 4, 6 in the $YY_8$-matrix (Figure 2) and in the multiplication table (Figure 5). All members of the $m_1$-sub-set of the basic elements $m_1$, $m_3$, $m_5$, $m_7$ and of their coordinates $x_1$, $x_3$, $x_5$, $x_7$ have the odd indexes 1, 3, 5, 7. And they occupy the columns with the odd numbers 1, 3, 5, 7 in the $YY_8$-matrix (Figure 2) and in the multiplication table (Figure 5).

By Pythagorean and Ancient Chinese traditions all even numbers are named "female" numbers or Yin-numbers, and all odd numbers are named "male" numbers or Yang-numbers. In accordance with these traditions one can name the elements $f_0$, $f_2$, $f_4$, $f_6$, $x_0$, $x_2$, $x_4$, $x_6$ with the even indexes as "female" or Yin-elements and the elements $m_1$, $m_3$, $m_5$, $m_7$, $x_1$, $x_3$, $x_5$, $x_7$ with the odd indexes as "male" elements or Yang-elements conditionally. By analogy one can name the columns with the even numerations 0, 2, 4, 6 (with the odd numerations 1, 3, 5, 7) as the female columns (the male columns). In this reason this octave algebra of the genetic code is named "the octave algebra with two quasi-real units" or the octave Yin-Yang-algebra (or the bisex algebra, or the even-odd-algebra). Such algebra gives new effective possibilities to model binary oppositions in biological objects at different levels, including triplets, amino acids, male and female gametal cells, male and female chromosomes, etc.

Each genetic triplet, which is disposed together with one of the female YY-coordinates $x_0$, $x_2$, $x_4$, $x_6$ in a mutual matrix cell, is named the female triplet or the Yin-triplet (Figure 3). The third position of all female triplets is occupied by the letter $\gamma$, which corresponds to the pyrimidine C or U/T. In this reason the female triplets can be named "pyrimidine triplets" as well. Each triplet, which is disposed together with one of the male YY-coordinates $x_1$, $x_3$, $x_5$, $x_7$ in a mutual matrix cell, is named the male triplet or the Yang-triplet. The third position of all male triplets is occupied by the letter $\delta$, which corresponds to the purine A or G. In this reason the male triplets can be named "purine triplets". In such algebraic way the whole set of 64 triplets is divided into two sub-sets of Yin-triplets (or female triplets) and Yang-triplets (or male triplets). We shall demonstrate later that the set of 20 amino acids is divided into the sub-sets of "female amino acids", "male amino acids" and "androgyne amino acids" from the this matrix viewpoint.

The multiplication table (Figure 5) is not symmetric one relative to the main diagonal; it corresponds to the non-commutative property of the Yin-Yang algebra. The expression (3) is the vector form of presentation of the genetic $YY_8$-number for the case of the genomatrix $P^{CAUG}_{123}{}^{(3)}$. It reminds the vector form of presentation of hypercomplex numbers $x_0*1+x_1*i_1+x_2*i_2+x_3*i_3+x_4*i_4+\ldots$ . But the significant difference exists between hypercomplex numbers and Yin-Yang-numbers. All cells of the main diagonal of multiplication tables for hypercomplex numbers are occupied always by the real unit only (with the signs "+" or "-"). On the contrary, all cells of the main diagonal of multiplication tables for $YY_8$-numbers are occupied by two quasi-real units $f_0$ and $m_1$ (with the signs "+" or "-") without the real unit at all (Figure 5). By their definition "hypercomplex numbers are the elements of the algebras with the real unit" [Mathematical encyclopedia, 1977]. Complex and hypercomplex numbers were constructed historically as generalizations of real numbers with the obligatory inclusion of the real unit in sets of their basic elements. It can be demonstrated easily that Yin-Yang algebras are the special generalization of the algebras of hypercomplex numbers. YY-numbers become the appropriate hypercomplex numbers in those cases when all their female (or male) coordinates are equal to zero. In other words, YY-numbers are the special generalization of hypercomplex numbers in the form of "double-hypercomplex" numbers. Traditional hypercomplex numbers can be represented as the "mono-sex" (Yin or Yang) half of appropriate YY-numbers. The algorithm of such generalization will be described later. We will denote Yin-Yang numbers by double letters (for example, YY) to distinguish them from traditional (complex and hypercomplex) numbers. In comparison with hypercomplex numbers, Yin-Yang numbers are the new category of numbers in the mathematical natural sciences in principle. In our opinion, knowledge of this category of numbers is necessary for deep understanding of biological phenomena, and, perhaps, it will be useful for mathematical natural sciences in the whole. Mathematical theory of YY-numbers gives new formal and conceptual apparatus to model phenomena of reproduction, self-organization and self-development in living nature.

The set of the basic elements of the $YY_8$-algebra forms a semi-group. Two squares are marked out by bold lines in the left upper corner of the multiplication table on Figure 5. The first two basic elements $f_0$ and $m_1$ are disposed in the smaller (2x2)-square of this table only. The greater (4x4)-square collects the four first basic elements $f_0$, $m_1$, $f_2$, $m_3$, which do not meet outside this square in the table also. These aspects say that sub-algebras $YY_2$ and $YY_4$ exist inside the algebra $YY_8$. We shall return to these sub-algebras in the Appendix A.2.

### 2.3 The structural analogies between the genomatrix $P^{CAUG}_{123}{}^{(3)}$ and the matrix $YY_8$

One should remind that the black cells of the genomatrix $P^{CAUG}_{123}{}^{(3)}$ contain the black NN-triplets, which encode the 8 high-degeneracy amino acids and the coding meaning of which do not depend on the letter on their third position (see details in [Petoukhov, arXiv:0802.3366; arXiv:0803.0888]). The set of the 8 high-degeneracy amino acids contains those amino acids, each of which is encoded by 4 triplets or more: Ala, Arg, Gly, Leu, Pro, Ser, Thr, Val [Petoukhov, 2005]. The white cells of the genomatrix $P^{CAUG}_{123}{}^{(3)}$ contain the white NN-triplets, the coding meaning of which depends on the letter on their third position and which encode the 12 low-degeneracy amino acids together with stop-signals. And the set of the 12 low-degeneracy amino acids contains those amino acids, each of which is encoded by 3 triplets or less: Asn, Asp, Cys, Gln, Glu, His, Ile, Lys, Met, Phe, Trp, Tyr.

The table on Figure 6 shows significant analogies and interrelations between the matrix $YY_8$ and the genomatrix $P^{CAUG}_{123}{}^{(3)}$ (Figures 1-3). Such structural coincidence of two matrices $YY_8$ and $P^{CAUG}_{123}{}^{(3)}$ permits to consider the octave algebra $YY_8$ as the adequate model of the structure of the genetic code. One can postulate such algebraic model and then deduce some peculiarities of the genetic code from this model.

| The octave Yin-Yang matrix $YY_8$ | The octave genomatrix $P^{CAUG}_{123}{}^{(3)}$ |
|---|---|
| This matrix possesses the binary mosaic of symmetrical character. It contains 32 YY-coordinates with the sign "+" and 32 YY-coordinates with the sign "-". | This genomatrix possesses the same binary mosaic. The black cells contain the high-degeneracy amino acids which are encoded by the 32 black NN-triplets. The white cells contain the low-degeneracy amino acids and the stop-signals which are encoded by the 32 white NN-triplets. |
| The enumerated matrix rows 0 and 1, 2 and 3, 4 and 5, 6 and 7 are equivalent to each other by a disposition of identical YY-coordinates. | The enumerated matrix rows 0 and 1, 2 and 3, 4 and 5, 6 and 7 are equivalent to each other by a disposition of identical amino acids. |
| The half of kinds of YY-coordinates ($x_0$, $x_1$, $x_2$, $x_3$) is presented in the quadrants along the main matrix diagonal only. The second half of kinds of YY-coordinates ($x_4$, $x_5$, $x_6$, $x_7$) is presented in the quadrants along the second coordinates only. | The half of kinds of amino acids is presented in the quadrants along the main matrix diagonal only (Ala, Arg, Asp, Gln, Glu, Gly, His, Leu, Pro, Val). The second half of kinds of amino acids is presented in the quadrants along the second diagonal only (Asn, Cys, Ile, Lys, Met, Phe, Ser, Thr, Trp, Tyr). |
| The YY-coordinates $x_0$, $x_2$, $x_4$, $x_6$ from the $f_0$-sub-set occupy the columns with even numbers 0, 2, 4, 6. The YY-coordinates $x_1$, $x_3$, $x_5$, $x_7$ from the $m_1$-sub-set occupy the columns with the odd numbers 1, 3, 5, 7. | The triplets with the pyrimidine C or U/T on their third position occupy the columns with the even numbers 0, 2, 4, 6. The triplets with the purine A or G on their third position occupy the columns with the odd numbers 1, 3, 5, 7. |

Figure 6. Examples of the conformity between the matrix $YY_8$ and the genomatrix $P^{CAUG}_{123}{}^{(3)}$.

The results of the comparison analysis in this table give the following answer to the question about mysterious principles of the degeneracy of the genetic code from the viewpoint of the proposed algebraic model. The matrix disposition of the 20 amino acids and the stop-signals is determined in some essential features by algebraic principles of the matrix disposition of the YY-coordinates. Moreover the disposition of the 32 black triplets and the high-degeneracy amino acids is determined by the disposition of the YY-coordinates with the sign "+"; the disposition of the 32 white triplets, the low-degeneracy amino acids and stop-signals is determined by the disposition of the YY-coordinates with the sign "-". One can remind here that the division of the set of the 20 amino acids into the two sub-sets of the 8 high-degeneracy amino acids and the 12 low-degeneracy amino acids is the invariant rule of all the dialects of the genetic code practically (see details in [Petoukhov, 2005]). The data of the table on Figure 6 do not exhaust the interconnections between the genetic code systems and the Yin-Yang matrices at all [Petoukhov, 2008b].

## 3 The Yin-Yang octave algebras and the permutations of positions in triplets

Our previous article [Petoukhov, arXiv:0803.0888] has presented the fact that the six possible variants of permutations of three positions in triplets (1-2-3, 2-3-1, 3-1-2, 1-3-2, 2-1-3, 3-2-1) generate the family of the six genomatrices $P^{CAUG}_{123}{}^{(3)}$, $P^{CAUG}_{231}{}^{(3)}$, $P^{CAUG}_{312}{}^{(3)}$, $P^{CAUG}_{132}{}^{(3)}$, $P^{CAUG}_{213}{}^{(3)}$, $P^{CAUG}_{321}{}^{(3)}$. For the considered case of the vertebrate mitochondria genetic code, all these genomatrices have symmetrologic mosaics of the code degeneracy. These data says that the degeneracy of the code has non-trivial connections with the position permutations in triplets.

Each triplet has its own YY-coordinate from the set $x_0, x_1, \ldots, x_7$ with the appropriate sign (Figure 3). The position permutations in triplets leads to new matrix dispositions of the triplets together with their coordinates $x_0, x_1, \ldots, x_7$. In such way new genomatrices $P^{CAUG}_{231}{}^{(3)}$, $P^{CAUG}_{312}{}^{(3)}$, $P^{CAUG}_{132}{}^{(3)}$, $P^{CAUG}_{213}{}^{(3)}$, $P^{CAUG}_{321}{}^{(3)}$ originate from the initial genomatrix $P^{CAUG}_{123}{}^{(3)}$. Algebraic properties of these matrices can be analyzed specially.

Above we have demonstrated that the initial genomatrix $P^{CAUG}_{123}{}^{(3)}$ of this permutation family possessed the interrelation with the $YY_8$-algebra. But the described permutation transformation of the genomatrix $P^{CAUG}_{123}{}^{(3)}$ into new genomatrices $P^{CAUG}_{231}{}^{(3)}$, $P^{CAUG}_{312}{}^{(3)}$, $P^{CAUG}_{132}{}^{(3)}$, $P^{CAUG}_{213}{}^{(3)}$, $P^{CAUG}_{321}{}^{(3)}$ can destroy this interrelation. For example the set of basic elements of each of these new genomatrices can be an unclosed set, and algebras do not originate in this case, or this set can be connected with algebras of quite other type. One can demonstrate in general case that arbitrary permutations of the columns and of the rows of the $P^{CAUG}_{231}{}^{(3)}$ lead in general case to new matrices, which possess unclosed sets of their basic elements. For instance, if the first and the second columns in the matrix $P^{CAUG}_{231}{}^{(3)}$ (or in the matrix $YY_8$ on Figure 2) interchange their places, the new matrix does not fit the YY-algebra at all. Generally speaking, a very little probability exists that these new genomatrices $P^{CAUG}_{231}{}^{(3)}$, $P^{CAUG}_{312}{}^{(3)}$, $P^{CAUG}_{132}{}^{(3)}$, $P^{CAUG}_{213}{}^{(3)}$, $P^{CAUG}_{321}{}^{(3)}$ fit $YY_8$-algebras also. But if this unexpected fact would be revealed, this fact will be the strong evidence of the deep interrelation between Yin-Yang algebras and the genetic code additionally.

Such unexpected fact was revealed by the author really: each of the genomatrices $P^{CAUG}_{231}{}^{(3)}$, $P^{CAUG}_{312}{}^{(3)}$, $P^{CAUG}_{132}{}^{(3)}$, $P^{CAUG}_{213}{}^{(3)}$, $P^{CAUG}_{321}{}^{(3)}$ fits their own $YY_8$-algebra. Each of these genomatrices (with the eight coordinates $x_0, x_1, \ldots, x_7$, which correspond to proper triplets) possesses its own set of the eight basic elements and its own multiplication table, which determines an octave algebra with two quasi-real units also. We shall mark these Yin-Yang algebras by the symbols $(YY_8)^{CAUG}_{231}$, $(YY_8)^{CAUG}_{312}$, etc. by analogy with the appropriate genomatrices $P^{CAUG}_{231}{}^{(3)}$, $P^{CAUG}_{312}{}^{(3)}$, etc. Each of these algebras possesses its own set of basic elements $f_0, m_1, f_2, m_3, f_4, m_5, f_6, m_7$. In other words, the matrix presentations of these basic elements differ from each other in the cases of the different algebras $(YY_8)^{CAUG}_{231}$, $(YY_8)^{CAUG}_{312}$, etc., though we utilize the same symbols for them here. And both quasi-real units have different forms of their matrix presentations in different Yin-Yang octave algebras also.

Figure 7 shows the example of the genomatrix $P^{CAUG}_{231}{}^{(3)}$ with the same coordinate from the set $x_0, x_1, \ldots, x_7$ for each triplet as in the genomatrix $P^{CAUG}_{123}{}^{(3)}$ on Figure 3. It can be checked that the (8*8)-matrix with such disposition of coordinates $x_0, x_1, \ldots, x_7$ is the matrix form of presentation of the Yin-Yang octave algebra $(YY_8)^{CAUG}_{231}$, the multiplication table of which is shown on Figure 8. The basic elements **f₀** and **m₄** occupy the main diagonal and play the role of the quasi-real units for this algebra.

| CCC | CAC | ACC | AAC | CCA | CAA | ACA | AAA |
|---|---|---|---|---|---|---|---|
| $x_0$ | $-x_2$ | $x_4$ | $-x_6$ | $x_1$ | $-x_3$ | $x_5$ | $-x_7$ |
| CUC | CGC | AUC | AGC | CUA | CGA | AUA | AGA |
| $x_2$ | $x_0$ | $-x_6$ | $-x_4$ | $x_3$ | $x_1$ | $-x_7$ | $-x_5$ |
| UCC | UAC | GCC | GAC | UCA | UAA | GCA | GAA |
| $x_4$ | $-x_6$ | $x_0$ | $-x_2$ | $x_5$ | $-x_7$ | $x_1$ | $-x_3$ |
| UUC | UGC | GUC | GGC | UUA | UGA | GUA | GGA |
| $-x_6$ | $-x_4$ | $x_2$ | $x_0$ | $-x_7$ | $-x_5$ | $x_3$ | $x_1$ |
| CCU | CAU | ACU | AAU | CCG | CAG | ACG | AAG |
| $x_0$ | $-x_2$ | $x_4$ | $-x_6$ | $x_1$ | $-x_3$ | $x_5$ | $-x_7$ |
| CUU | CGU | AUU | AGU | CUG | CGG | AUG | AGG |
| $x_2$ | $x_0$ | $-x_6$ | $-x_4$ | $x_3$ | $x_1$ | $-x_7$ | $-x_5$ |
| UCU | UAU | GCU | GAU | UCG | UAG | GCG | GAG |
| $x_4$ | $-x_6$ | $x_0$ | $-x_2$ | $x_5$ | $-x_7$ | $x_1$ | $-x_3$ |
| UUU | UGU | GUU | GGU | UUG | UGG | GUG | GGG |
| $-x_6$ | $-x_4$ | $x_2$ | $x_0$ | $-x_7$ | $-x_5$ | $x_3$ | $x_1$ |

Figure 7. The disposition of coordinates $x_0, x_1, \ldots, x_7$ in the genomatrix $P^{CAUG}_{231}{}^{(3)}$ reproduced from the article [Petoukhov, arXiv:0803.0888].

|  | $f_0$ | $f_1$ | $f_2$ | $f_3$ | $m_4$ | $m_5$ | $m_6$ | $m_7$ |
|---|---|---|---|---|---|---|---|---|
| $f_0$ | $f_0$ | $f_1$ | $f_2$ | $f_3$ | $m_4$ | $m_5$ | $m_6$ | $m_7$ |
| $f_1$ | $f_1$ | $-f_0$ | $-f_3$ | $f_2$ | $m_5$ | $-m_4$ | $-m_7$ | $m_6$ |
| $f_2$ | $f_2$ | $f_3$ | $f_0$ | $f_1$ | $m_6$ | $m_7$ | $m_4$ | $m_5$ |
| $f_3$ | $f_3$ | $-f_2$ | $-f_1$ | $f_0$ | $m_7$ | $-m_6$ | $-m_5$ | $m_4$ |
| $m_4$ | $f_0$ | $f_1$ | $f_2$ | $f_3$ | $m_4$ | $m_5$ | $m_6$ | $m_7$ |
| $m_5$ | $f_1$ | $-f_0$ | $-f_3$ | $f_2$ | $m_5$ | $-m_4$ | $-m_7$ | $m_6$ |
| $m_6$ | $f_2$ | $f_3$ | $f_0$ | $f_1$ | $m_6$ | $m_7$ | $m_4$ | $m_5$ |
| $m_7$ | $f_3$ | $-f_2$ | $-f_1$ | $f_0$ | $m_7$ | $-m_6$ | $-m_5$ | $m_4$ |

Figure 8. The multiplication table of the basic elements of the octave Yin-Yang-algebra $(YY_8)^{CAUG}_{231}$ for the genomatrix $P^{CAUG}_{231}{}^{(3)}$ on Figure 7. The elements **f₀** and **m₄** are the quasi-real units in this algebra.

|  | $f_0$ | $f_1$ | $m_2$ | $m_3$ | $f_4$ | $f_5$ | $m_6$ | $m_7$ |
|---|---|---|---|---|---|---|---|---|
| $f_0$ | $f_0$ | $f_1$ | $m_2$ | $m_3$ | $f_4$ | $f_5$ | $m_6$ | $m_7$ |
| $f_1$ | $f_1$ | $f_0$ | $m_3$ | $m_2$ | $f_5$ | $f_4$ | $m_7$ | $m_6$ |
| $m_2$ | $f_0$ | $f_1$ | $m_2$ | $m_3$ | $f_4$ | $f_5$ | $m_6$ | $m_7$ |
| $m_3$ | $f_1$ | $f_0$ | $m_3$ | $m_2$ | $f_5$ | $f_4$ | $m_7$ | $m_6$ |
| $f_4$ | $f_4$ | $-f_5$ | $m_6$ | $-m_7$ | $-f_0$ | $f_1$ | $-m_2$ | $m_3$ |
| $f_5$ | $f_5$ | $-f_4$ | $m_7$ | $-m_6$ | $-f_1$ | $f_0$ | $-m_3$ | $m_2$ |
| $m_6$ | $f_4$ | $-f_5$ | $m_6$ | $-m_7$ | $-f_0$ | $f_1$ | $-m_2$ | $m_3$ |
| $m_7$ | $f_5$ | $-f_4$ | $m_7$ | $-m_6$ | $-f_1$ | $f_0$ | $-m_3$ | $m_2$ |

Figure 9. The multiplication table of the basic elements of the octave Yin-Yang-algebra $(YY_8)^{CAUG}_{312}$ for the genomatrix $P^{CAUG}_{312}{}^{(3)}$. The elements **f₀** and **m₂** are the quasi-real units in this algebra.

|       | $f_0$ | $f_1$ | $m_2$ | $m_3$ | $f_4$ | $f_5$ | $m_6$ | $m_7$ |
|-------|-------|-------|-------|-------|-------|-------|-------|-------|
| $f_0$ | $f_0$ | $f_1$ | $m_2$ | $m_3$ | $f_4$ | $f_5$ | $m_6$ | $m_7$ |
| $f_1$ | $f_1$ | $-f_0$ | $m_3$ | $-m_2$ | $-f_5$ | $f_4$ | $-m_7$ | $m_6$ |
| $m_2$ | $f_0$ | $f_1$ | $m_2$ | $m_3$ | $f_4$ | $f_5$ | $m_6$ | $m_7$ |
| $m_3$ | $f_1$ | $-f_0$ | $m_3$ | $-m_2$ | $-f_5$ | $f_4$ | $-m_7$ | $m_6$ |
| $f_4$ | $f_4$ | $f_5$ | $m_6$ | $m_7$ | $f_0$ | $f_1$ | $m_2$ | $m_3$ |
| $f_5$ | $f_5$ | $-f_4$ | $m_7$ | $-m_6$ | $-f_1$ | $f_0$ | $-m_3$ | $m_2$ |
| $m_6$ | $f_4$ | $f_5$ | $m_6$ | $m_7$ | $f_0$ | $f_1$ | $m_2$ | $m_3$ |
| $m_7$ | $f_5$ | $-f_4$ | $m_7$ | $-m_6$ | $-f_1$ | $f_0$ | $-m_3$ | $m_2$ |

Figure 10. The multiplication table of the basic elements of the octave Yin-Yang-algebra $(YY_8)^{CAUG}_{132}$ for the genomatrix $P^{CAUG}_{132}{}^{(3)}$. The elements $f_0$ and $m_2$ are the quasi-real units in this algebra.

|       | $f_0$ | $m_1$ | $f_2$ | $m_3$ | $f_4$ | $m_5$ | $f_6$ | $m_7$ |
|-------|-------|-------|-------|-------|-------|-------|-------|-------|
| $f_0$ | $f_0$ | $m_1$ | $f_2$ | $m_3$ | $f_4$ | $m_5$ | $f_6$ | $m_7$ |
| $m_1$ | $f_0$ | $m_1$ | $f_2$ | $m_3$ | $f_4$ | $m_5$ | $f_6$ | $m_7$ |
| $f_2$ | $f_2$ | $m_3$ | $f_0$ | $m_1$ | $f_6$ | $m_7$ | $f_4$ | $m_5$ |
| $m_3$ | $f_2$ | $m_3$ | $f_0$ | $m_1$ | $f_6$ | $m_7$ | $f_4$ | $m_5$ |
| $m_5$ | $f_4$ | $m_5$ | $-f_6$ | $-m_7$ | $-f_0$ | $-m_1$ | $f_2$ | $m_3$ |
| $m_5$ | $f_4$ | $m_5$ | $-f_6$ | $-m_7$ | $-f_0$ | $-m_1$ | $f_2$ | $m_3$ |
| $f_6$ | $f_6$ | $m_7$ | $-f_4$ | $-m_5$ | $-f_2$ | $-m_3$ | $f_0$ | $m_1$ |
| $m_7$ | $f_6$ | $m_7$ | $-f_4$ | $-m_5$ | $-f_2$ | $-m_3$ | $f_0$ | $m_1$ |

Figure 11. The multiplication table of the basic elements of the octave Yin-Yang-algebra $(YY_8)^{CAUG}_{213}$ for the genomatrix $P^{CAUG}_{213}{}^{(3)}$. The elements $f_0$ and $m_1$ are the quasi-real units in this algebra.

|       | $f_0$ | $f_1$ | $f_2$ | $f_3$ | $m_4$ | $m_5$ | $m_6$ | $m_7$ |
|-------|-------|-------|-------|-------|-------|-------|-------|-------|
| $f_0$ | $f_0$ | $f_1$ | $f_2$ | $f_3$ | $m_4$ | $m_5$ | $m_6$ | $m_7$ |
| $f_1$ | $f_1$ | $f_0$ | $f_3$ | $f_2$ | $m_5$ | $m_4$ | $m_7$ | $m_6$ |
| $f_2$ | $f_2$ | $-f_3$ | $-f_0$ | $f_1$ | $m_6$ | $-m_7$ | $-m_4$ | $m_5$ |
| $f_3$ | $f_3$ | $-f_2$ | $-f_1$ | $f_0$ | $m_7$ | $-m_6$ | $-m_5$ | $m_4$ |
| $m_4$ | $f_0$ | $f_1$ | $f_2$ | $f_3$ | $m_4$ | $m_5$ | $m_6$ | $m_7$ |
| $m_5$ | $f_1$ | $f_0$ | $f_3$ | $f_2$ | $m_5$ | $m_4$ | $m_7$ | $m_6$ |
| $m_6$ | $f_2$ | $-f_3$ | $-f_0$ | $f_1$ | $m_6$ | $-m_7$ | $-m_4$ | $m_5$ |
| $m_7$ | $f_3$ | $-f_2$ | $-f_1$ | $f_0$ | $m_7$ | $-m_6$ | $-m_5$ | $m_4$ |

Figure 12. The multiplication table of the basic elements of the Yin-Yang-algebra $(YY_8)^{CAUG}_{321}$ for the genomatrix $P^{CAUG}_{321}{}^{(3)}$. The elements $f_0$ and $m_4$ are the quasi-real units in this algebra.

Figures 9-12 demonstrate the multiplications tables of the basic elements of the Yin-Yang algebras for the other genomatrices $P^{CAUG}_{312}{}^{(3)}$, $P^{CAUG}_{132}{}^{(3)}$, $P^{CAUG}_{213}{}^{(3)}$, $P^{CAUG}_{321}{}^{(3)}$ from the family of the six permutation genomatrices described in [Petoukhov, arXiv:0803.0888].

Taking into account the multiplication tables on Figures 6, 8-12 the proper $YY_8$-numbers in the vector form of their presentation have the following expressions:

$$(YY_8)^{CAUG}_{123} = x_0*f_0+x_1*m_1+x_2*f_2+x_3*m_3+x_4*f_4+x_5*m_5+x_6*f_6+x_7*m_7$$
$$(YY_8)^{CAUG}_{231} = x_0*f_0+x_1*f_1+x_2*f_2+x_3*f_2+x_4*m_4+x_5*m_5+x_6*m_6+x_7*m_7$$
$$(YY_8)^{CAUG}_{312} = x_0*f_0+x_1*f_1+x_2*m_2+x_3*m_3+x_4*f_4+x_5*f_5+x_6*m_6+x_7*m_7$$
$$(YY_8)^{CAUG}_{132} = x_0*f_0+x_1*f_1+x_2*m_2+x_3*m_3+x_4*f_4+x_5*f_5+x_6*m_6+x_7*m_7$$
$$(YY_8)^{CAUG}_{213} = x_0*f_0+x_1*m_1+x_2*f_2+x_3*m_3+x_4*f_4+x_5*m_5+x_6*f_6+x_7*m_7$$
$$(YY_8)^{CAUG}_{321} = x_0*f_0+x_1*f_1+x_2*f_2+x_3*f_3+x_4*m_4+x_5*m_5+x_6*m_6+x_7*m_7 \quad (4)$$

All these Yin-Yang matrices have secret connections with Hadamard matrices: when all their coordinates are equal to the real unit 1 ($x_0=x_1=\ldots=x_7=1$) and when the change of signs of some components of the matrices takes place by means of the U-algorithm described in the article [Petoukhov, arXiv:0802.3366], then all these Yin-Yang octave matrices become the Hadamard matrices. In necessary cases biological computers of organisms can transform these Yin-Yang matrices into the Hadamard matrices to operate with systems of orthogonal vectors. One can add that for the case when all their coordinates are equal to 1 ($x_0=x_1=\ldots=x_7=1$), all these six Yin-Yang matrices $(YY_8)^{CAUG}_{123}$, $(YY_8)^{CAUG}_{231}$, …, $(YY_8)^{CAUG}_{321}$ possess the property of their tetra-reproduction which was described in the work [Petoukhov, arXiv:0803.0888] and which reminds the tetra-reproduction of gametal cells in the biological process of meiosis.

One can mention two facts else. The complementary triplets (codon and anti-codon) play essential role in the genetic code systems. One can replace each codon in the genomatrices $P^{CAUG}_{123}$, $P^{CAUG}_{231}$, $P^{CAUG}_{312}$, $P^{CAUG}_{132}$, $P^{CAUG}_{213}$, $P^{CAUG}_{321}$ by its anti-codon. The new six genomatrices appear in this case. Have they any connection with Yin-Yang algebras? We have investigated this question with the positive result. The multiplication tables for the basic elements of Yin-Yang matrices, connected with these new genomatrices, are identical to the multiplication tables for the initial genomatrices. In other words, the "complementary" transformations of the genomatrices $P^{CAUG}_{123}$, $P^{CAUG}_{231}$, $P^{CAUG}_{312}$, $P^{CAUG}_{132}$, $P^{CAUG}_{213}$, $P^{CAUG}_{321}$ change the matrix forms of the initial $YY_8$-numbers only but do not change the Yin-Yang algebras of the genomatrices. But if we consider the transposed matrices, which are received from the matrices $(YY_8)^{CAUG}_{123}$, $(YY_8)^{CAUG}_{231}$, etc., they correspond to new Yin-Yang octave algebras.

**4 The genetic Yin-Yang octaves as "double genoquaternions"**

We shall name any numbers with 8 items $x_0*\mathbf{i_0}+x_1*\mathbf{i_1}+\ldots x_7*\mathbf{i_7}$ by the name "octaves" independently of multiplication tables of their basic elements. We shall name numbers with 4 items $x_0*\mathbf{i_0}+x_1*\mathbf{i_1}+x_2*\mathbf{i_2}+x_3*\mathbf{i_3}$ by the name "quaternions" independently of multiplication tables of their basic elements (quaternions by Hamilton are the special case of quaternions). Let us analyze the expression (3) of the genetic octave $YY_8$ together with its multiplication table (Figure 5). If all male coordinates are equal to zero ($m_1=m_3=m_5=m_7$), this genetic octave $YY_8$ becomes the genetic quaternion $g_f$:

$$g_f = x_0*\mathbf{f_0} + x_2*\mathbf{f_2} + x_4*\mathbf{f_4} + x_6*\mathbf{f_6} \qquad (5)$$

The proper multiplication table for this quaternion is shown on Figure 13 (on the left side). This table is received from the multiplication table for the algebra $YY_8$ (Figure 5) by nullification (by excision) of the columns and rows, which have the male basic elements. Taking into account that the basic element $\mathbf{f_0}$ possesses the multiplication properties of the real unit relative to all female basic elements, one can rewrite the expression (5) in the following form:

$$g_f = x_0*\mathbf{1} + x_2*\mathbf{f_2} + x_4*\mathbf{f_4} + x_6*\mathbf{f_6} \qquad (6)$$

If all female coordinates are equal to zero ($f_0=f_2=f_4=f_6$), this genetic octave $YY_8$ becomes the genetic quaternion $g_m$:

$$g_m = x_1*\mathbf{m_1} + x_3*\mathbf{m_3} + x_5*\mathbf{m_5} + x_7*\mathbf{m_7} \qquad (7)$$

The appropriate multiplication table for this quaternion is shown on Figure 13 (on the right side). Taking into account that the basic element $\mathbf{m_1}$ possesses the multiplication properties of the real unit relative to all male basic elements, one can rewrite the expression (7) in the following form:

$$g_m = x_1*\mathbf{1} + x_3*\mathbf{m_3} + x_5*\mathbf{m_5} + x_7*\mathbf{m_7} \qquad (8)$$

|   | $f_0$ | $f_2$ | $f_4$ | $f_6$ |
|---|---|---|---|---|
| $f_0$ | $f_0$ | $f_2$ | $f_4$ | $f_6$ |
| $f_2$ | $f_2$ | $-f_0$ | $-f_6$ | $f_4$ |
| $f_4$ | $f_4$ | $f_6$ | $f_0$ | $f_2$ |
| $f_6$ | $f_6$ | $-f_4$ | $-f_2$ | $f_0$ |

|   | $m_1$ | $m_3$ | $m_5$ | $m_7$ |
|---|---|---|---|---|
| $m_1$ | $m_1$ | $m_3$ | $m_5$ | $m_7$ |
| $m_3$ | $m_3$ | $-m_1$ | $-m_7$ | $m_5$ |
| $m_5$ | $m_5$ | $m_7$ | $m_1$ | $m_3$ |
| $m_7$ | $m_7$ | $-m_5$ | $-m_3$ | $m_1$ |

Figure 13. The multiplication tables for the genetic quaternions $g_f$ (on the left side) and $g_m$ (on the right side).

The quaternions $g_f$ and $g_m$ are similar to each other. They can be expressed in the following general form, the multiplication table of which is shown on Figure 14 (on the right side):

$$g = y_0*\mathbf{1} + y_1*\mathbf{i_1} + y_2*\mathbf{i_2} + y_3*\mathbf{i_3} \qquad (9)$$

Figure 14 shows the comparison between the multiplication tables for quaternions by Hamilton (on the left side) and for these genetic quaternions g (or briefly "genoquaternions").

|   | 1 | $i_1$ | $i_2$ | $i_3$ |
|---|---|---|---|---|
| 1 | 1 | $i_1$ | $i_2$ | $i_3$ |
| $i_1$ | $i_1$ | $-1$ | $i_3$ | $-i_2$ |
| $i_2$ | $i_2$ | $-i_3$ | $-1$ | $i_1$ |
| $i_3$ | $i_3$ | $i_2$ | $-i_1$ | $-1$ |

|   | 1 | $i_1$ | $i_2$ | $i_3$ |
|---|---|---|---|---|
| 1 | 1 | $i_1$ | $i_2$ | $i_3$ |
| $i_1$ | $i_1$ | $-1$ | $-i_3$ | $i_2$ |
| $i_2$ | $i_2$ | $i_3$ | 1 | $i_1$ |
| $i_3$ | $i_3$ | $-i_2$ | $-i_1$ | 1 |

Figure 14. The multiplication tables for quaternions by Hamilton (on the left side) and for genoquaternions (on the right side).

| Quaternions by Hamilton | Genoquaternions |
|---|---|
| $q = x_0*\mathbf{1} + x_1*\mathbf{i_1} + x_2*\mathbf{i_2} + x_3*\mathbf{i_3}$ | $g = x_0*\mathbf{1} + x_1*\mathbf{i_1} + x_2*\mathbf{i_2} + x_3*\mathbf{i_3}$ |
| $(q_1*q_2)*q3 = q_1*(q_2*q3)$ | $(g_1*g_2)*g3 = g_1*(g_2*g3)$ |
| Conjugate quaternion $q_s = x_0*\mathbf{1} - x_1*\mathbf{i_1} - x_2*\mathbf{i_2} - x_3*\mathbf{i_3}$ | Conjugate genoquaternion $g_s = x_0*\mathbf{1} - x_1*\mathbf{i_1} - x_2*\mathbf{i_2} - x_3*\mathbf{i_3}$ |
| To the norm of quaternions: $\|q\|^2 = q*q_s = q_s*q = x_0^2 + x_1^2 + x_2^2 + x_3^2$ | To the norm of genoquaternions: $\|g\|^2 = g*g_s = g_s*g = x_0^2 + x_1^2 - x_2^2 - x_3^2$ |
| The inverse quaternion exists: $q^{-1} = q_s/\|q\|^2$ | The inverse genoquaternion exists: $g^{-1} = g_s/\|g\|^2$ |
| $(q_1 + q_2)_s = (q_1)_s + (q_2)_s$ | $(g_1 + g_2)_s = (g_1)_s + (g_2)_s$ |
| $(q_1*q_2)_s = (q_2)_s * (q_1)_s$ | $(g_1*g_2)_s = (g_2)_s * (g_1)_s$ |

Figure 15. The comparison of some properties between the systems of quaternions by Hamilton (on the left side) and of genoquaternions (on the right side).

The system of quaternions by Hamilton has many useful properties and applications in mathematics and physics. The author has received the essential result that the system of genoquaternions possesses many analogical properties, which permits to think about its useful applications in bioinformatics, mathematical biology, etc. For example, the numeric system of genoquaternions is the system with the operation of division and it possesses the associative property, the notions of the "norm of genoquaternion" and of the "inverse genoquaternion", etc. Figure 15 demonstrates some analogies between both types of quaternions.

Taking into account the expressions (5-9), one can name the genetic octave $x_0*\mathbf{i_0}+x_1*\mathbf{i_1}+...x_7*\mathbf{i_7}$ (with its individual multiplication table) as "the double genoquaternion". This name generates heuristic associations with the famous name "the double spiral" of DNA.

## 5 The comparison between the classical vector calculation and the genovector calculation

The theory of quaternions by W.Hamilton possesses many useful results and applications. Let us remind about one of them, which concerns the beautiful connection between these quaternions $q = x_0*\mathbf{1}+x_1*\mathbf{i_1}+x_2*\mathbf{i_2}+ x_3*\mathbf{i_3}$ and the classical vector calculation developed by J.Gibbs. One can take two vectors $\underline{a}$ and $\underline{b}$, which belong to the plane $(\mathbf{i_v}, \mathbf{i_w})$, where v<w, v = 1, 2; w = 2, 3; $\underline{a} = a_1*\mathbf{i_v} + a_2*\mathbf{i_w}$, $\underline{b} = b_1*\mathbf{i_1} + b_2*\mathbf{i_2}$. These vectors can be presented in the following usual form:

$$\underline{a} = |\underline{a}|*(\mathbf{i_v}*\cos \alpha + \mathbf{i_w}*\sin \alpha), \qquad \underline{b} = |\underline{b}|*(\mathbf{i_v}*\cos \beta + \mathbf{i_w}*\sin \beta), \qquad (10)$$

where α and β are appropriate angles between these vectors and the axises $\mathbf{i_v}$ and $\mathbf{i_w}$ in the orthogonal system of the basic vectors $(\mathbf{i_1}, \mathbf{i_2}, \mathbf{i_3})$. If we multiply together these vectors as Hamilton's quaternions in accordance with the multiplication table on Figure 14 (the left side), the following equation arises:

$$\underline{a}*\underline{b} = - |\underline{a}|*|\underline{b}|*\cos(\alpha - \beta) + |\underline{a}|*|\underline{b}|*\sin(\alpha - \beta)*\mathbf{i_{vw}}, \qquad (11)$$

where $\mathbf{i_{vw}}$ is the third basic vector, which is orthogonal to the basic vectors $\mathbf{i_v}$ and $\mathbf{i_m}$ .

The expression (11) shows that the quaternion multiplication of two vectors contains two parts: the scalar part and the vector part. The scalar part $|\underline{a}|*|\underline{b}|*\cos(\alpha - \beta)$ is famous under the name "the scalar product" and the vector part $\underline{a}|*|\underline{b}|*\sin(\alpha - \beta)*\mathbf{i_3}$ is famous under the name "the vector product" in the classical vector calculation. This vector calculation is utilized widely in mechanics to describe movements of hard bodies in our physical space, etc. Mechanics of bodies in the usual physical space fits this vector calculation. From the viewpoint of this vector calculation, the space is isotropic one because the expression (11) with its scalar and vector parts is the same for each of pairs of vectors, which belong to the planes $(\mathbf{i_1}, \mathbf{i_2})$, $(\mathbf{i_1}, \mathbf{i_3})$, $(\mathbf{i_2}, \mathbf{i_3})$, and the scalar products and the vectors product possess the analogical forms for all three cases of the planes.

But what results arise, if we multiply together the vectors $\underline{a}$ and $\underline{b}$ (10) as genoquaternions in accordance with their multiplication table (Figure 14, on the right side)? Let us consider the following three cases, each of which contains a scalar part and a vector part in the final expressions (12), (13), (14), but in different forms.

<u>The case 1</u>. The vectors $\underline{a}$ and $\underline{b}$ belong to the plane $(\mathbf{i_1}, \mathbf{i_2})$. They can be expressed in the following form: $\underline{a} = |\underline{a}|*(\mathbf{i_1}*\cos \alpha + \mathbf{i_2}*\sin \alpha)$, $\underline{b} = |\underline{b}|*(\mathbf{i_1}*\cos \beta + \mathbf{i_2}*\sin \beta)$. If we multiply together these vectors as genoquaternions (Figure 14, in the right side), the result arises:

$$a*b = |a|*|b|*(\mathbf{i_1}*\cos \alpha+\mathbf{i_2}*\sin \alpha)*(\mathbf{i_1}*\cos \beta+\mathbf{i_2}*\sin \beta) = - |a|*|b|*\cos(\alpha+\beta)+|a|*|b|*\sin(\alpha-\beta)*\mathbf{i_3} \quad (12)$$

The expression (12) of the genovector calculation differs from the expression (11) of the classical vector calculation in the scalar part only (by the value $\cos(\alpha+\beta)$).

<u>The case 2</u>. The vectors **_a_** and **_b_** belong to the plane ($i_1$, $i_3$):   **_a_** = |**_a_**|*($i_1$*cos α + $i_3$*sin α), **_b_** = |**_b_**|*($i_1$*cos β + $i_3$*sin β).  The product of these two vectors as genoquaternions gives the following result:

**_a_**\***_b_** = |**_a_**|\*|**_b_**|\*($i_1$\*cos α+$i_3$\*sin α)\*($i_1$\*cos β+$i_3$\*sin β) = -|**_a_**|\*|**_b_**|\*cos(α+β) - |**_a_**|\*|**_b_**|\*sin(α-β)\*$i_2$   (13)

The expression (13) of the genovector calculation differs from the classical expression (11) in the scalar part (by the value cos(α+β)) and in the vector part (by the opposite sign).

<u>The case 3</u>. The vectors **_a_** and **_b_** belong to the plane ($i_2$, $i_3$):   **_a_** = |**_a_**|*($i_2$*cos α + $i_3$*sin α), **_b_** = |**_b_**|*($i_2$*cos β + $i_3$*sin β).  The product of these two vectors as genoquaternions gives

**_a_**\***_b_** = |**_a_**|\*|**_b_**|\*($i_2$\*cos α+$i_3$\*sin α)\*($i_2$\*cos β+$i_3$\*sin β) =+|**_a_**|\*|**_b_**|\*cos(α-β) - |**_a_**|\*|**_b_**|\*sin(α-β)\*$i_1$   (14)

The expression (14) of the genovector calculation differs from the classical expression (11) by the opposite sign in the scalar part and in the vector part.

We name vectors, which are considered as qenoquaternions (with applications of the rules of genoquaternion operations to them), as "genovectors". It is obvious that the genovector calculation fits the case of an anisotropic space because the results of multiplication of arbitrary vectors **_a_** and **_b_** depend on the plane, to which these vectors belong. The spaces of biological phenomena of morphogenesis, growth, etc. have anisotropic characters also. Since the genovector calculation was developed by the author from the genetic code features, it seems that this calculation (and its generalization for the system of Yin-Yang genooctaves) can be adequate to model anisotropic processes in biological spaces including processes of bioinformatics and of biological morphogenesis on different levels of each united organism.

Many mathematical formalisms and notions, which were received in the theory of quaternions by Hamilton and which were utilized in many scientific branches, have their analogies in the theory of genoquaternions (and in the theory of genetic tetrions described below). Details of these analogies and their possible applications in different scientific branches will be described in our next publications.

## 6   The parametric reduction of the genetic octave Yin-Yang-algebra to the 4-dimensional algebra of tetrions. About the pre-code and its alphabet

The previous paragraphs have considered the numeric system $YY_8$ = $x_0$\***f_0**+$x_1$\***m_1**+$x_2$\***f_2**+$x_3$\***m_3**+$x_4$\***f_4**+$x_5$\***m_5**+$x_6$\***f_6**+$x_7$\***m_7** (3) with 8 arbitrary coordinates $x_0$, $x_1$, …, $x_7$. But in accordance with the matrix on Figure 3 all these 8 coordinates are expressed by means of 4 parameters α, β, γ, δ:

$$x_0 = ααγ;\quad x_1 = ααδ;\quad x_2 = αβγ;\quad x_3 = αβδ;\quad x_4 = βαγ;\quad x_5 = βαδ;\quad x_6 = ββγ;\quad x_7 = ββδ \quad (15)$$

Hence these 8 coordinates are not independent on each other and they are interconnected by the following expressions:

$$x_1 = x_0*δ/γ;\quad x_3 = x_2*δ/γ;\quad x_5 = x_4*δ/γ;\quad x_7 = x_6*δ/γ \quad (16)$$

One can see from the expression (15) that the coordinates belong to the female (male) type if they have the symbol γ (δ correspondingly) on their third position. The expressions (16) show the existence of the pairs of "complementary" male and female coordinates which differ by the coefficient δ/γ only: $x_1$ and $x_0$; $x_3$ and $x_2$; $x_5$ and $x_4$; $x_7$ and $x_6$. These interconnections of coordinates lead to the particular form of the octave number $YY_8$, where female coordinates $x_0$, $x_2$, $x_4$, $x_6$ exist only (another possible form has the male coordinates $x_1$, $x_3$, $x_5$, $x_7$ only):

$$T = x_0*(f_0+\delta/\gamma*m_1) + x_2*(f_2+\delta/\gamma*m_3) + x_4*(f_4+\delta/\gamma*m_5) + x_6*(f_6+\delta/\gamma*m_7) =$$

$$=\alpha\alpha\gamma * \begin{vmatrix} 1 & \delta/\gamma & 0 & 0 & 0 & 0 & 0 & 0 \\ 1 & \delta/\gamma & 0 & 0 & 0 & 0 & 0 & 0 \\ 0 & 0 & 1 & \delta/\gamma & 0 & 0 & 0 & 0 \\ 0 & 0 & 1 & \delta/\gamma & 0 & 0 & 0 & 0 \\ 0 & 0 & 0 & 0 & 1 & \delta/\gamma & 0 & 0 \\ 0 & 0 & 0 & 0 & 1 & \delta/\gamma & 0 & 0 \\ 0 & 0 & 0 & 0 & 0 & 0 & 1 & \delta/\gamma \\ 0 & 0 & 0 & 0 & 0 & 0 & 1 & \delta/\gamma \end{vmatrix} + \alpha\beta\gamma * \begin{vmatrix} 0 & 0 & 1 & \delta/\gamma & 0 & 0 & 0 & 0 \\ 0 & 0 & 1 & \delta/\gamma & 0 & 0 & 0 & 0 \\ -1 & -\delta/\gamma & 0 & 0 & 0 & 0 & 0 & 0 \\ -1 & -\delta/\gamma & 0 & 0 & 0 & 0 & 0 & 0 \\ 0 & 0 & 0 & 0 & 0 & 0 & 1 & \delta/\gamma \\ 0 & 0 & 0 & 0 & 0 & 0 & 1 & \delta/\gamma \\ 0 & 0 & 0 & 0 & -1 & -\delta/\gamma & 0 & 0 \\ 0 & 0 & 0 & 0 & -1 & -\delta/\gamma & 0 & 0 \end{vmatrix}$$

$$+\beta\alpha\gamma * \begin{vmatrix} 0 & 0 & 0 & 0 & -1 & -\delta/\gamma & 0 & 0 \\ 0 & 0 & 0 & 0 & -1 & -\delta/\gamma & 0 & 0 \\ 0 & 0 & 0 & 0 & 0 & 0 & 1 & \delta/\gamma \\ 0 & 0 & 0 & 0 & 0 & 0 & 1 & \delta/\gamma \\ -1 & -\delta/\gamma & 0 & 0 & 0 & 0 & 0 & 0 \\ -1 & -\delta/\gamma & 0 & 0 & 0 & 0 & 0 & 0 \\ 0 & 0 & 1 & \delta/\gamma & 0 & 0 & 0 & 0 \\ 0 & 0 & 1 & \delta/\gamma & 0 & 0 & 0 & 0 \end{vmatrix} + \beta\beta\gamma * \begin{vmatrix} 0 & 0 & 0 & 0 & 0 & 0 & -1 & -\delta/\gamma \\ 0 & 0 & 0 & 0 & 0 & 0 & -1 & -\delta/\gamma \\ 0 & 0 & 0 & 0 & -1 & -\delta/\gamma & 0 & 0 \\ 0 & 0 & 0 & 0 & -1 & -\delta/\gamma & 0 & 0 \\ 0 & 0 & -1 & -\delta/\gamma & 0 & 0 & 0 & 0 \\ 0 & 0 & -1 & -\delta/\gamma & 0 & 0 & 0 & 0 \\ -1 & -\delta/\gamma & 0 & 0 & 0 & 0 & 0 & 0 \\ -1 & -\delta/\gamma & 0 & 0 & 0 & 0 & 0 & 0 \end{vmatrix}$$ (17)

Each of these four matrices $(f_0+\delta/\gamma*m_1)$, $(f_2+\delta/\gamma*m_3)$, $(f_4+\delta/\gamma*m_5)$, $(f_6+\delta/\gamma*m_7)$ is constructed by means of fusion of male and female basic matrices of the complementary pair into united object. It is interesting that these four matrices form their own closed set relative to multiplication. Figure 16 shows the table of multiplication of these matrices.

| | $f_0+\delta/\gamma*m_1$ | $f_2+\delta/\gamma*m_3$ | $f_4+\delta/\gamma*m_5$ | $f_6+\delta/\gamma*m_7$ |
|---|---|---|---|---|
| $f_0+\delta/\gamma*m_1$ | $(1+\delta/\gamma)*(f_0+\delta/\gamma*m_1)$ | $(1+\delta/\gamma)*(f_2+\delta/\gamma*m_3)$ | $(1+\delta/\gamma)*(f_4+\delta/\gamma*m_5)$ | $(1+\delta/\gamma)*(f_6+\delta/\gamma*m_7)$ |
| $f_2+\delta/\gamma*m_3$ | $(1+\delta/\gamma)*(f_2+\delta/\gamma*m_3)$ | $-(1+\delta/\gamma)*(f_0+\delta/\gamma*m_1)$ | $-(1+\delta/\gamma)*(f_6+\delta/\gamma*m_7)$ | $(1+\delta/\gamma)*(f_4+\delta/\gamma*m_5)$ |
| $f_4+\delta/\gamma*m_5$ | $(1+\delta/\gamma)*(f_4+\delta/\gamma*m_5)$ | $(1+\delta/\gamma)*(f_6+\delta/\gamma*m_7)$ | $(1+\delta/\gamma)*(f_0+\delta/\gamma*m_1)$ | $(1+\delta/\gamma)*(f_2+\delta/\gamma*m_3)$ |
| $f_6+\delta/\gamma*m_7$ | $(1+\delta/\gamma)*(f_6+\delta/\gamma*m_7)$ | $-(1+\delta/\gamma)*(f_4+\delta/\gamma*m_5)$ | $-(1+\delta/\gamma)*(f_2+\delta/\gamma*m_3)$ | $(1+\delta/\gamma)*(f_0+\delta/\gamma*m_1)$ |

Figure 16. The table of multiplication of the matrices $(f_0+\delta/\gamma*m_1)$, $(f_2+\delta/\gamma*m_3)$, $(f_4+\delta/\gamma*m_5)$, $(f_6+\delta/\gamma*m_7)$, which are basic elements of the genetic tetrions.

In this reason the expression (17) with all possible values of real numbers α, β, γ, δ represents the new system of 4-dimensional numbers, which are named "genetic tetrions" (or genotetrions) to distinguish them from 4-dimensional hypercomplex numbers called "quaternions" traditionally (including genoquaternions described above). If quaternions and other hypercomplex numbers have the real unit among their basic elements, tetrions have not the real unit among their basic elements at all. The first basic element $(f_0+\delta/\gamma*m_1)$ of the tetrions (17) is the matrix presentation of the real number $(1+\delta/\gamma)$. This basic element possesses the commutative property relative to all these basic elements. The first item $x_0*(f_0+\delta/\gamma*m_1)$ is considered as the scalar part of tetrions, and other three items $x_2*(f_2+\delta/\gamma*m_3) + x_4*(f_4+\delta/\gamma*m_5) + x_6*(f_6+\delta/\gamma*m_7)$ are considered as the vector part of tetrions.

The square of any basic element of the tetrions T is equal to $(1+\delta/\gamma)*(f_0+\delta/\gamma*m_1)$ with the sign "+" or "-". This peculiarity is demonstrated on Figure 16 in cells (marked by bold borders) along the main diagonal. So instead of the real unit, tetrions have the real number (1+v), where "v" is the real number, which is equal to $\delta/\gamma$ in the case of the genetic tetrions T (17). One can consider such tetrions as the special generalization of appropriate hypercomplex numbers by means of utilizing any kind of real numbers in the role of their first basic element instead of utilizing the real unit in this role in the case of traditional hypercomplex numbers. (Concerning to such form of generalization of hypercomplex numbers, for example one can write the matrix presentation of similar generalization of complex numbers: $y_0*[1, v, 0, 0; 1, v, 0, 0; 0, 0, 1, v; 0, 0, 1, v]+y_1*[0, 0, 1, v; 0, 0, 1, v; -1, -v, 0, 0; -1, -v, 0, 0]$).

The system of tetrions T (17) possesses the commutative and associative properties. It is the system with operation of division from the left side and from the right side (by analogy with the division in the system of quaternions). By definition the conjugate tetrion $T_S$ is presented by the expression:

$$T_S = x_0*(f_0+\delta/\gamma*m_1) - x_2*(f_2+\delta/\gamma*m_3) - x_4*(f_4+\delta/\gamma*m_5) - x_6*(f_6+\delta/\gamma*m_7) \qquad (18)$$

The following expressions for two tetrions $T_1$ and $T_2$ hold true:

$$(T_1 + T_2)_S = (T_1)_S + (T_2)_S ; \qquad (T_1*T_2)_S = (T_2)_S * (T_1)_S \qquad (19)$$

The square of the module of tetrions:

$$|T|^2 = T*T_S = T_S*T = (1+\delta/\gamma)*(x_0^2+x_2^2-x_4^2-x_6^2) = (1+\delta/\gamma)*[(\alpha\alpha\gamma)^2+(\alpha\beta\gamma)^2-(\beta\alpha\gamma)^2-(\beta\beta\gamma)^2] \qquad (20)$$

The inverse genotetrion exists: $T^{-1} = T_S/|T|^2$. It permits to define the operation of division traditionally by means of multiplication by the inverse genotetrion. One can see that these properties of the genetic tetrions are similar to the properties of genoquaternions considered above (Figure 15) and that the genotetrion's and genoquaternion's tables of multiplication are similar to each other by the disposition of the signs "+" and "-" (Figures 14 and 16).

The system of genetic tetrions leads to special kind of vector calculation. By analogy with the expressions (12-14) for genoquaternions, one can receive the similar expressions (21-23) of vector calculation for genotetrions. Let us analyse the multiplication of two vectors *a* and *b* (10) as tetrions in accordance with the multiplication table (16) in the same three cases which were described in the section 5. In the result we receive the following expressions (21-23).

<u>The case 1</u>. The vectors *a* and *b* belong to the plane of the basic vectors ($f_2+\delta/\gamma*m_3$, $f_4+\delta/\gamma*m_5$). Then

$$a*b = - |a|*|b|*(1+\delta/\gamma)^2*\cos(\alpha+\beta) + |a|*|b|*\sin(\alpha-\beta)*(1+\delta/\gamma)*(f_6+\delta/\gamma*m_7). \qquad (21)$$

<u>The case 2</u>. The vectors *a* and *b* belong to the plane ($f_2+\delta/\gamma*m_3$, $f_6+\delta/\gamma*m_7$). Then

$$a*b = - |a|*|b|*(1+\delta/\gamma)^2*\cos(\alpha+\beta) - |a|*|b|*\sin(\alpha-\beta)*(1+\delta/\gamma)*(f_4+\delta/\gamma*m_5). \qquad (22)$$

<u>The case 3</u>. The vectors *a* and *b* belong to the plane ($f_2+\delta/\gamma*m_3$, $f_6+\delta/\gamma*m_7$). Then

$$a*b = +|a|*|b|*(1+\delta/\gamma)^2*\cos(\alpha-\beta) - |a|*|b|*\sin(\alpha-\beta)*(1+\delta/\gamma)*(f_2+\delta/\gamma*m_3). \qquad (23)$$

It is obvious that the vector calculation of genetic tetrions fits the case of an anisotropic space because the results of multiplication of arbitrary vectors *a* and *b* depend on the plane, to which these vectors belong. Can the scalar and vector parts of genetic tetrions be considered correspondingly as the time coordinate and the space coordinates in the theory of the genetic space-time? This and other interesting questions are under investigation now and their answers should be published later.

In the described approach, the genetic code is presented as the replica of the tetrions in their matrix form. It permits to consider the algebra of genetic tetrions as a candidate for the role of the mathematical system of genetic preceding code (the "pre-code" or the more fundamental code) relative to the genetic code. Really, from a traditional viewpoint, a code is an aggregate of symbols which corresponds to elements of information. In our algebraic case, the speech is about the matrix system, symbols of which can be confronted with triplets and with other elements of the genetic code. In other words, the genetic code can be encoded itself by symbols of elements of the tetrion numerical system. Such tetrion pre-code has its own pre-code alphabet, which consists of the four letters $\alpha, \beta, \gamma, \delta$ in contrast to the usual genetic alphabet A, C, G, U/T. This set of the letters $\alpha, \beta, \gamma, \delta$ can be named the alphabet of genetic algebras or the algebraic alphabet of the genetic code as well. A revealing such tetrion pre-code as new numeric system can help with sorting, ordering and deeper understanding of the genetic information. It can help also to develop new effective methods of processing and transfer of information in many applied problems. Mathematical features of such pre-code can explain evolutionary features of the genetic code.

One should emphasize that not only the (8x8)-matrix $YY_8$ (Figures 2 and 3), but each of its (4x4)-quadrants and each of its (2x2)-subquadrants define its own special algebras, if we take into account the coordinates $x_0, x_1, \ldots, x_7$ and the algebraic alphabet $\alpha, \beta, \gamma, \delta$. It means that the genetic code is an ensemble of special multidimensional algebras from such matrix viewpoint. Details of this statement will be published by the author separately.

### 7 About genetic mechanics

In the beginning of the XIX century the following opinion existed: the world possesses the single real geometry (Euclidean geometry) and the single arithmetic. But this opinion was neglected after the discovery of non-Euclidean geometries and of quaternions by Hamilton. The science understood that different natural systems can possess their own individual geometries and their own individual algebras (see this theme in the book [Kline, 1980]). The example of Hamilton, who has wasted 10 years in his attempts to solve the task of description of transformations of 3D space by means of 3-dimensional algebras without a success, is the very demonstrative one. This example says that if a scientist does not guess right what type of algebras is adequate for the natural system, which is investigated by him, he can waste many years without result by analogy with Hamilton. One can add that geometrical and physical-geometrical properties of separate natural systems (including laws of conservations, theories of oscillations and waves, theories of potentials and fields, etc.) can depend on the type of algebras which are adequate for them.

The fact, that the genetic code has led us to the algebra of genetic tetrions (which is the special case of the genetic octave Yin-Yang-algebra), testifies in favor of the importance of this algebra for each united organism. It seems that many difficulties of modern science to understand genetic and biological systems are determined by approaches to these systems from the viewpoint of non-adequate algebras, which were developed formerly for other systems at all. In particular, the classical vector calculation, which plays the role of the important tool in classical mechanics and which fit geometrical properties of our physical space, can be inappropriate for important biological phenomena.

We put forward the hypothesis that a very special mechanics of biogenetic systems exists, which is connected with the vector calculation of genetic tetrions and with their generalization in the form of Yin-Yang octaves. We name it "genetic mechanics" because of its relation with the genetic code. Modern biomechanics is the set of applications of classical mechanics to model properties of living substances. In our opinion, such traditional biomechanics are not adequate to many biological phenomena and it will be replaced in many aspects by the genetic mechanics in future. We think that living substance lives in its own biological space which has specific algebraic and geometric properties.

The hypothesis about a non-Euclidean geometry of living nature exists long ago [Vernadsky, 1965] but without a concrete definition of the type of such geometry. And how one can construct such geometry if biological organisms – bacteria, birds, fishes, plants, etc. - differ from each other so significantly in their morphogenetic and many other features? The discovery of the genetic code, the basic elements of which are general for all biological organisms, has permitted to hope that such geometric and algebraic tasks can be solved by means of investigation of genetic code structures. Some results of such investigation are presented in this article.

It happens frequently, that mathematicians construct a new beautiful abstract mathematics and then they search for opportunities of its application in different areas of natural sciences. On the contrary, in our case the phenomenology of the genetic code has led the author unexpectedly to the new mathematics of tetrions and Yin-Yang-octaves. And we investigate formal features of this mathematics on the second stage only. The genetic code is the result of a gigantic experiment of the nature. This molecular code bears the imprint of a great set of known and unknown laws of the nature. In this connection, algebraic features of genetic structures are very essential to guess right a perspective direction of development of algebraic bases of mathematical natural sciences in future. In our opinion, the tetrion algebra, the Yin-Yang-algebra and their geometries can be useful not only for biology, but also for other fields of mathematical natural sciences and for applied sciences (signals processing, mathematical economy, etc.). For example, they permit to develop new algorithms and methods of digital signal processing.

## 8   The genetic code as the multidimensional number and the idea by Pythagoras

The notion "number" is the main notion of mathematics. In accordance with the famous thesis, the complexity of civilization is reflected in complexity of numbers which are utilized by the civilization [Mathematics in the modern world, 1964]. "Number - one of the most fundamental concepts not only in mathematics, but also in all natural sciences. Perhaps, it is the more primary concept than such global categories, as time, space, substance or a field" [Hypercomplex numbers in geometry and in physics, 2004].

After establishment of real numbers in the history of development of the notion "number", complex numbers $x_0+i*x_1$ have appeared. These 2-dimensional numbers have played the role of the magic tool for development of theories and calculations in the field of problems of heat, light, sounds, fluctuations, elasticity, gravitation, magnetism, electricity, current of liquids, the quantum-mechanical phenomena. It seems that modern atomic stations, airplanes, rockets and many other things would not exist without knowledge of complex numbers because the appropriate physical theories are based on these numbers. C.Gauss, J.Argand and C. Wessel have demonstrated that a plane with its properties fits 2-dimensional complex numbers. W.Hamilton has proved that the properties of our 3-dimensional physical space fit mathematical properties of the special quaternions. Hamilton's quaternions have played the significant role in the history of mathematical natural sciences as well. For example, the classical vector calculation is deduced from the theory of these quaternions. The author supposes that genetic tetrions, genoquaternions and Yin-Yang octaves ("double genoquaternions"), which are prompted by the genetic code and which are new kind of generalized numbers, can play a significant role in appropriate scientific fields in future also.

Pythagoras has formulated the famous idea: "All things are numbers". This idea had a great influence on mankind. B.Russel wrote that he did not know any other person who has influenced with such power on other people in the field of thought [Russel, 1945]. In this relation the world does not know the more fundamental scientific idea than the idea by Pythagoras (it should be mentioned that the notion "number" was perfected in science after Pythagoras in the direction of generalized numbers such as hypercomplex numbers). We reveal the fact that the genetic code with its degeneracy properties fits genetic tetrions (which are the special case of Yin-Yang octaves or "double genoquaternions") by analogy with the fact that the 3D physical space fits Hamilton's quaternions. One can say in this relation that the genetic code is the tetrion number (in a certain sense). Our results give additional materials to the great idea by Pythagoras.

"Number" is the main notion of mathematics. In the result of our investigations of the genetic code as the bases of bioinformatics, we find ourselves in the field of fundamentals of mathematics and mathematical natural sciences unexpectedly. It has many mathematical and heuristic consequences.

## 9 About the tetrion algebra of the genetic code as the algebra of operators

This brief paragraph pays attention to one of the most perspective directions of understanding of the possible biological meanings of the genetic tetrion algebra and of the Yin-Yang-algebras. It is known that mathematics has deals not only with algebras of numbers but with algebras of operators also (see historical remarks in the book [Kliene,1980, Chapter VIII]). G.Boole has published in 1854 year his brilliant work about investigations of laws of thinking. He has proposed the Boole's algebra of logics (or logical operators). Boole tried to construct such operator algebra which would reflect basic properties of human thinking. The Boole's algebra plays a great role in the modern science because of its connections with many scientific branches: mathematical logic, the problem of artificial intelligence, computer technologies, bases of theory of probability, etc.

In our opinion, the described algebra of tetrions and Yin-Yang-algebra of the genetic code can be considered not only as the algebras of the numeric systems but as the algebra of proper logical operators of genetic systems also. This direction of thoughts can lead us to deeper understanding of logic of biological systems including an advanced variant of the idea by Boole (and by some other scientists) about development of algebraic theory of laws of thinking. The author plans to publish a possible variant of such genetic algebras of logical operators later. One can add that biological organisms have famous possibilities to utilize the same structures in multi-purpose destinations. The genetic algebras can be utilized by biological organisms in different purposes also.

## 10 What is life from the viewpoint of algebra? The problem of algebraization of bioinformatics

Taking into account the great meaning of the genetic code for biological organisms, the described discovery of algebraic properties of the genetic code gives the basis for investigation of biological organizations from the algebraic viewpoint. Modern algebra is the wide branch of mathematics. It possesses many own theorems, applications of which to genetic systems can give new vision in the field of theoretical biology. It is essential also that algebra plays the great role in the modern theory of information encoding and of signal processing. All these facts provoke the high interest to the question: what is life from the viewpoint of algebra?

This question exists now in parallel with the old question from the famous book by E.Schrodinger: what is life from the viewpoint of physics? One can add that attempts are known in modern theoretical physics to reveal information bases of physics; in these attempts information principles are considered as the most fundamental.

Here one can mention as well the known problem of geometrization of physics, that is the problem of creation and interpretation of physical theories in a form of theories of invariants of groups of transformations. Such general approach to different physical theories was very fruitful. One can hope that the problem of algebraization of bioinformatics (and of biology, which is connected closely with bioinformatics), that is understanding bioinformation phenomena from the viewpoint of algebras of the genetic code, will be useful also.

### Appendix A: Some additional mathematical peculiarities of the genetic octave Yin-Yang-algebra $YY_8$

#### A.1 Interrelations between hypercomplex numbers and Yin-Yang numbers

As it was mentioned above, Yin-Yang numbers (YY-numbers) can be considered as the generalization of hypercomplex numbers. Each of $2^{n-1}$-dimensional hypercomplex numbers can be transformed into the $2^n$-dimensional YY-number by a special algorithm. An inverse application of this algorithm to a $2^n$-dimensional YY-number generates the appropriate $2^{n-1}$-dimensional hypercomplex number. According to this algorithm, if we have a $(2^n \times 2^n)$-matrix, which represents a $2^n$-dimensional hypercomplex number, we should replace each component of this matrix by the (2x2)-matrix $[x_к \ x_{к+1}; \ x_к \ x_{к+1}]$. By this algorithm we have the tetra-reproduction of matrix components which reminds the tetra-reproduction of gametal cells in the result of meiosis. In such reason this algorithm has the conditional name "the meiosis algorithm".

For example, if we have the (2x2)-matrix of the presentation of complex numbers, this meiosis algorithm transforms it into (4x4)-matrix of the presentation of 4-dimensional "Yin-Yang-complex" numbers $KK_4$, which fit the special multiplication table of the appropriate $YY_4$-algebra (Figure 17). Really, according to this algorithm, each component $x_0$ and $x_1$ of the initial matrix is replaced by the (2x2)-matrix of the mentioned type: $x_0=[y_0 \ y_1; \ y_0 \ y_1]$, $x_1=[y_2 \ y_3; \ y_2 \ y_3]$. In the result we have YY-complex numbers $КК_4 = y_0*f_0+y_1*m_1+y_2*f_2+y_3*m_3$, where $f_0$ and $m_1$ are the female and male quasi-real units; $f_2$ и $m_3$ are the female and male imaginary units with the properties $(f_2)^2 = -f_0$, $(m_3)^2 = -m_1$.

| | | | | | | $f_0$ | $m_1$ | $f_2$ | $m_3$ |
|---|---|---|---|---|---|---|---|---|---|
| | | $y_0$ | $y_1$ | $-y_2$ | $-y_3$ | $f_0$ | $f_0$ | $m_1$ | $f_2$ | $m_3$ |
| $x_0$ | $-x_1$ | $y_0$ | $y_1$ | $-y_2$ | $-y_3$ | $m_1$ | $f_0$ | $m_1$ | $f_2$ | $m_3$ |
| $x_1$ | $x_0$ | $y_2$ | $y_3$ | $y_0$ | $y_1$ | $f_2$ | $f_2$ | $m_3$ | $-f_0$ | $-m_1$ |
| | | $y_2$ | $y_3$ | $y_0$ | $y_1$ | $m_3$ | $f_2$ | $m_3$ | $-f_0$ | $-m_1$ |

Figure 17. The matrix forms of presentation of complex numbers (on the left side) and YY-complex numbers (in the middle). On the right side: the multiplication table for basic elements of the YY-complex numbers.

By inverse application of this algorithm, one can receive a 4-dimensional hypercomplex number from the genetic YY-number $YY_8$. The YY-matrix $YY_8$ (Figure 2) contains 4 kinds of (2x2)-sub-quadrants, each of which has one of pairs of coordinates: $x_0$ and $x_1$; $x_2$ and $x_3$; $x_4$ and $x_5$; $x_6$ and $x_7$. One can replace each such sub-quadrant by a separate coordinate: $[x_0 \ x_1; \ x_0 \ x_1] = y_0$;

[$x_2$ $x_3$; $x_2$ $x_3$] = $y_1$; [$x_4$ $x_5$; $x_4$ $x_5$] = $y_2$; [$x_6$ $x_7$; $x_6$ $x_7$] = $y_3$. In the result the (4x4)-matrix Q is appeared, which represents the genoquaternion $Q = y_0*\mathbf{1} + y_1*\mathbf{i_1} + y_2*\mathbf{i_2} + y_3*\mathbf{i_3}$, which was considered above and which has $\mathbf{i_1}^2 = -1$, $\mathbf{i_2}^2 = \mathbf{i_3}^2 = +1$. Figure 18 shows the matrix Q and the multiplication table for this genoquaternion. The genoquaternion Q reminds coquaternions (or split-quaternions, or para-quaternions, or hyperbolic quaternions), introduced by J.Cockle in 1849 year (http://en.Qikipedia.org/Qiki/Coquaternion), but their multiplication tables have differences. We name the number Q "genoquaternion of the first type". (A genoquaternion of the second type is produced by the special permutation of columns of the matrix Q, which is connected with the permutation of positions in genetic duplets [Petoukhov, 2008b, p.203]).

$$Q = \begin{bmatrix} y_0 & -y_1 & y_2 & -y_3 \\ y_1 & y_0 & -y_3 & -y_2 \\ y_2 & -y_3 & y_0 & -y_1 \\ -y_3 & -y_2 & y_1 & y_0 \end{bmatrix}$$

|   | 1 | $i_1$ | $i_2$ | $i_3$ |
|---|---|---|---|---|
| 1 | 1 | $i_1$ | $i_2$ | $i_3$ |
| $i_1$ | $i_1$ | -1 | -$i_3$ | $i_2$ |
| $i_2$ | $i_2$ | $i_3$ | 1 | $i_1$ |
| $i_3$ | $i_3$ | -$i_2$ | -$i_1$ | 1 |

Figure 18. The matrix form of presentation of the hypercomplex number Q (on the left side); its multiplication table (on the right side).

### A.2 About the sub-algebras $YY_2$ and $YY_4$ of the genetic Yin-Yang octave algebra $YY_8$

Two squares are marked out by bold lines in the left top corner of the multiplication table on Figure 5 for the case of the genomatrix $P^{CAUG}_{123}$. The first of these squares with its size (2x2) is the multiplication table of the basic elements of the 2-dimensional Yin-Yang algebra $YY_2$. Figure 19 shows two matrix forms of presentation of appropriate Yin-Yang numbers $YY_2$. One of these forms [$z_0$ $z_1$; $z_0$ $z_1$] coincides with the structure of each (2x2)-sub-quadrant of the genomatrices on Figures 1-3 in relation of the disposition of the YY-coordinates $x_0, x_1,…, x_7$ and amino acids with stop-signals. It testifies that the algebra $YY_2$ participates in the structural organization of the genetic code.

The second square with its size (4x4) on Figure 5 is the multiplication table of the 4-dimensional Yin-Yang algebra $YY_4$. The appropriate Yin-Yang numbers $YY_4$ possess the following vector form: $YY_4 = z_0*\mathbf{f_0}+z_1*\mathbf{m_1}+z_2*\mathbf{f_2}+z_3*\mathbf{m_3}$ and these numbers coincide with the Yin-Yang generalization of complex numbers (Figure 17).

$$\begin{vmatrix} z_0 & z_1 \\ z_0 & z_1 \end{vmatrix} ; \begin{vmatrix} z_0 & -z_1 \\ -z_0 & z_1 \end{vmatrix} ;$$

|   | $1^F_0$ | $1^M_1$ |
|---|---|---|
| $1^F_0$ | $1^F_0$ | $1^M_1$ |
| $1^M_1$ | $1^F_0$ | $1^M_1$ |

Figure 19. Two matrix forms of a presentation of the 2-dimensional numbers $YY_2$ (on the left side); the multiplication table of the basic elements of the Yin-Yang algebra $YY_2$.

### A.3 About the geometrical interpretation of the 2-dimensional Yin-Yang numbers $YY_2$

It is known that complex numbers have been widely recognized only after finding of their geometrical interpretation on the geometric plane of complex variables. This plane was named "Gauss-Argand plane" according to names of the mathematicians who have introduced such plane. Whether it is possible to offer a substantial geometrical interpretation of the 2-dimensional Yin-Yang numbers $YY_2$? Yes, it is possible. For this purpose we introduce the plane of Yin-Yang variables (or YY-plane). It is an ordinary plane with the Yin-Yang system of Cartesian coordinates. This Yin-Yang system (or YY-system) has the coordinate axes **f** and **m**, which play a role of female and male axes. By analogy with the case of complex numbers, each 2-dimensional YY-number is denoted on this YY-plane by the point or by the vector. A product XX*ZZ of two Yin-Yang vectors, where **XX**= $x_0$*$f_0$ + $x_1$*$m_1$ and **ZZ**= $z_0$*$f_0$ + $z_1$*$m_1$, possesses a geometric sense on such plane. Really, the result of non-commutative multiplication of such two YY-vectors is equal to the second vector with the scale coefficient, which is equal to the sum of coordinates of the first vector (Figure 20, on the left side). The same first vector-factor at multiplication with all other vectors of the plane or of a geometric figure leads to their identical scaling (Figure 20, on the right side).

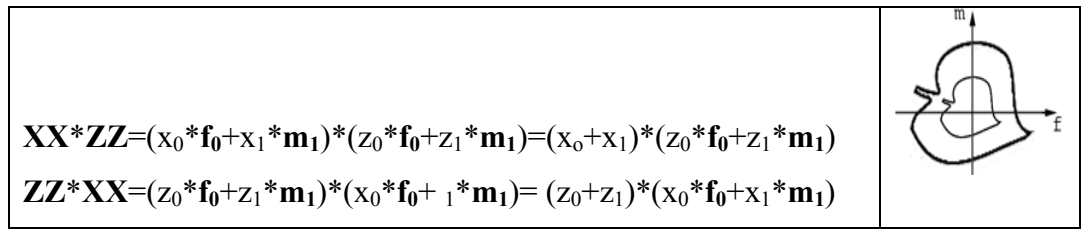

**XX*ZZ**=($x_0$*$f_0$+$x_1$*$m_1$)*($z_0$*$f_0$+$z_1$*$m_1$)=($x_o$+$x_1$)*($z_0$*$f_0$+$z_1$*$m_1$)

**ZZ*XX**=($z_0$*$f_0$+$z_1$*$m_1$)*($x_0$*$f_0$+ $_1$*$m_1$)= ($z_0$+$z_1$)*($x_0$*$f_0$+$x_1$*$m_1$)

Figure 20. The non-commutative multiplication of two Yin-Yang vectors (on the left side). A scaling of a geometric figure on the Yin-Yang plane (on the right side).

It associates with the known biological phenomenon of volumetric growth of the living bodies, observed at the most different lines and branches of biological evolution. Biological bodies are capable to the mysterious volumetric growth, occurring in the cooperative way in all volume of the body or of its growing part. It is one of sharp differences of living bodies from crystals with their surface growth occurring due to a local addition of new portions of substance on the surface of the crystal. In this connection, the Yin-Yang geometry is one of candidates for the role of the geometry of biological volumetric growth. Other details of the Yin-Yang geometry will be published later.

### A.4 About algebras with many quasi-real units (polysex algebras)

It should be mentioned that the names "bisex algebra", "bisex geometry", "bisex numbers", etc. can be utilized as the synonyms of the names "Yin-Yang algebra", "Yin-Yang geometry", "Yin-Yang numbers". In some cases the utilization of these names can be more comfortable but it depends on situations. For example it is comfortable in the question about algebras with many quasi-real units. Such algebras can be named "polysex algebras" (or "n-sex algebras").

One can denote that bisex algebras are a particular case of n-sex algebras, each of which possesses a set of their basic elements with "n" quasi-real units but without the real unit. Figure 21 shows the simplest example of 3-sex numbers $x_0$*$i_0$+$x_1$*$i_1$+$x_2$*$i_2$ (in the matrix form of their presentation), which contain three quasi-real numbers only. The basic elements $i_0$, $i_1$, $i_2$ of these 3-sex numbers have their matrix forms: $i_0$=[1 0 0; 1 0 0; 1 0 0], $i_1$=[0 1 0; 0 1 0; 0 1 0], $i_2$=[0 0 1; 0 0 1; 0 0 1]. Their multiplication table is shown on Figure 21.

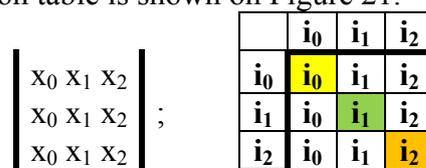

Figure 21. The matrix form of presentation of 3-sex numbers and the multiplication table of their basic elements.

Bisex numbers and trisex numbers can be considered as numeric analogies of the famous symbols Yin-Yang and tomoe (Figure 22). Details about the Japanese tomoe symbol are given at the site http://altreligion.about.com/library/glossary/symbols/bldefstomoe.htm .

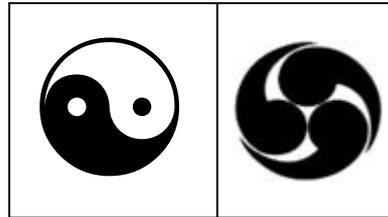

Figure 22. The symbol Yin-Yang and the symbol tomoe.

Multiplication of two 3-sexes gives the result which is similar to the described case of bisexes: the result is the 3-sex, which is equal to the second factor increased by the sum of coordinates of the first factor (Figure 23). The geometry of 3-sexes is a candidate to play the role of the geometry of the volumetric biological growth in the case of 3D-space (by analogy with the geometry of bisexes in the case of a plane).

$$\begin{vmatrix} x_0, x_1, x_3 \\ x_0, x_1, x_2 \\ x_0, x_1, x_2 \end{vmatrix} * \begin{vmatrix} y_0, y_1, y_3 \\ y_0, y_1, y_2 \\ y_0, y_1, y_2 \end{vmatrix} = (x_0+x_1+x_2)* \begin{vmatrix} y_0, y_1, y_3 \\ y_0, y_1, y_2 \\ y_0, y_1, y_2 \end{vmatrix}$$

Figure 23. Multiplication of two 3-sexes.

Figure 24 shows another example of polysexes: the matrix form of 8-dimensional 4-sexes $x_0*i_0+x_1*i_1+x_2*i_2+x_3*i_3+x_4*i_4+x_5*i_5+x_6*i_6+x_7*i_7$, which have 4 quasi-real units $i_0, i_1, i_2, i_3$ and which have its own imaginary unit for each of these quasi-real units: $i_4^2 = -i_0$; $i_5^2 = -i_1$; $i_6^2 = -i_2$; $i_7^2 = -i_3$.

$$\begin{vmatrix} x_0 & x_1 & x_2 & x_3 & -x_4 & -x_5 & -x_6 & -x_7 \\ x_0 & x_1 & x_2 & x_3 & -x_4 & -x_5 & -x_6 & -x_7 \\ x_0 & x_1 & x_2 & x_3 & -x_4 & -x_5 & -x_6 & -x_7 \\ x_0 & x_1 & x_2 & x_3 & -x_4 & -x_5 & -x_6 & -x_7 \\ x_4 & x_5 & x_6 & x_7 & x_0 & x_1 & x_2 & x_3 \\ x_4 & x_5 & x_6 & x_7 & x_0 & x_1 & x_2 & x_3 \\ x_4 & x_5 & x_6 & x_7 & x_0 & x_1 & x_2 & x_3 \\ x_4 & x_5 & x_6 & x_7 & x_0 & x_1 & x_2 & x_3 \end{vmatrix}$$

|       | $i_0$ | $i_1$ | $i_2$ | $i_3$ | $i_4$ | $i_5$ | $i_6$ | $i_7$ |
|-------|-------|-------|-------|-------|-------|-------|-------|-------|
| $i_0$ | $i_0$ | $i_1$ | $i_2$ | $i_3$ | $i_4$ | $i_5$ | $i_6$ | $i_7$ |
| $i_1$ | $i_0$ | $i_1$ | $i_2$ | $i_3$ | $i_4$ | $i_5$ | $i_6$ | $i_7$ |
| $i_2$ | $i_0$ | $i_1$ | $i_2$ | $i_3$ | $i_4$ | $i_5$ | $i_6$ | $i_7$ |
| $i_3$ | $i_0$ | $i_1$ | $i_2$ | $i_3$ | $i_4$ | $i_5$ | $i_6$ | $i_7$ |
| $i_4$ | $i_4$ | $i_5$ | $i_6$ | $i_7$ | $-i_0$ | $-i_1$ | $-i_2$ | $-i_3$ |
| $i_5$ | $i_4$ | $i_5$ | $i_6$ | $i_7$ | $-i_0$ | $-i_1$ | $-i_2$ | $-i_3$ |
| $i_6$ | $i_4$ | $i_5$ | $i_6$ | $i_7$ | $-i_0$ | $-i_1$ | $-i_2$ | $-i_3$ |
| $i_7$ | $i_4$ | $i_5$ | $i_6$ | $i_7$ | $-i_0$ | $-i_1$ | $-i_2$ | $-i_3$ |

Figure 24. The matrix form of presentation of 8-dimensional 4-sexes (the upper matrix) and their multiplication table (the lower table)

Descriptions of other mathematical properties of Yin-Yang algebras and their interrelations with the bioinformation systems will be continued in our next publications. YY-numbers permit to develop new class of mathematical models of self-reproduction systems. The author investigates bisex generalizations of known physical equations additionally to find new results with a physical sense from there (it is the mathematical fact that known physical equations can be received from appropriate bisex equations by passage to the limit in values of appropriate bisex coordinates). The algebraic theory of the genetic code can say many useful and unexpected things about an origin of the genetic code and about laws of living substances.

**Appendix B: About the notion "algebra"**

The notion "algebra" has two main senses. According to the first sense, which is famous more widely, the algebra is the whole section of mathematics about mathematical operations with mathematical symbols. According to the second sense, which is utilized in this article, algebra is a mathematical object with certain properties or, better to say, arithmetic of multidimensional numbers.

By definition in the frame of this second sense, algebra A with its dimension "n" over a field P is a set of expressions $x_0*i_0+x_1*i_1+x_2*i_2+\ldots+x_{n-1}*i_{n-1}$ (where $x_0, x_1,\ldots, x_{n-1}$ belong P; $i_0, i_1, \ldots i_{n-1}$ are basic elements of vectors, which fit such expressions). This set is provided with the operation of multiplication by elements "k" from the field P to determine the formula $k*(x_0*i_0 + x_1*i_1 +x_2*i_2 +\ldots+ x_{n-1}*i_{n-1}) = k*x_0*i_0 + k*x_1*i_1 + k*x_2*i_2 +\ldots+ k*x_{n-1}*i_{n-1}$. This set is provided with the following operation of addition also: $(x_0*i_0+x_1*i_1+x_2*i_2+\ldots+x_{n-1}*i_{n-1}) + (y_0*i_0+y_1*i_1+y_2*i_2+\ldots+y_{n-1}*i_{n-1}) = (x_0+y_0)*i_0 + (x_1+y_1)*i_1 + \ldots + (x_{n-1}+y_{n-1})*i_{n-1}$. This set is provided with the operation of multiplication between symbols $i_r$ else, which is given by a specific multiplication table $i_r*i_v = w_{rv,0}*i_0 + w_{rv,1}*i_1 +\ldots w_{rv,n-1}*i_2$. This multiplication table is utilized to find the result of multiplications $(x_0*i_0+x_1*i_1+x_2*i_2+\ldots+x_{n-1}*i_{n-1})*(y_0*i_0+y_1*i_1+y_2*i_2+ \ldots+y_{n-1}*i_{n-1})$. Any algebra is defined completely by its multiplication table that is by a certain set of numbers $w_{rv,u}$. These numbers do not subordinate to any conditions, and any such set of numbers defines the certain algebra.

**Acknowledgments**: Described researches were made by the author in the frame of a long-term cooperation between Russian and Hungarian Academies of Sciences and in the frame of programs of "International Society of Symmetry in Bioinformatics" (USA, http://polaris.nova.edu/MST/ISSB) and of "International Symmetry Association" (Hungary, http://symmetry.hu/). The author is grateful to Frolov K.V., Darvas G., Ne'eman Y., He M., Bakhtiarov K.I., Kassandrov V.V., Smolianinov V.V., Vladimirov Y.S. for their support.